\PassOptionsToPackage{usenames,dvipsnames}{xcolor}
\documentclass[sigconf, nonacm, pdfa]{acmart}
\usepackage{graphicx}
\usepackage{microtype}
\usepackage{epstopdf}
\usepackage{flushend}
\usepackage{balance}
\usepackage{colortbl}
\usepackage{xcolor}
\usepackage{listings}
\usepackage{booktabs}
\usepackage{nicefrac}
\usepackage{paralist}
\usepackage{caption}
\usepackage{subcaption}
\usepackage{soul}
\usepackage{subfiles}
\usepackage{multirow}
\usepackage{tabularx}
\usepackage{verbatim}
\usepackage{graphicx} 
\usepackage{pdfpages}

\usepackage[a-2b]{pdfx}
\usepackage{balance}
\usepackage{textcomp}
\usepackage{enumitem}
\usepackage{xspace}
\usepackage{appendix}
\usepackage{siunitx}

\definecolor{Blue}{rgb}{0.01, 0.28, 0.8}
\definecolor{comment-red}{rgb}{1,0,0}
\definecolor{OliveGreen}{rgb}{0,0.6,0}

\newcommand{\name}{{DEX}\xspace}
\newcommand{\sherman}{{Sherman}\xspace}
\newcommand{\smart}{{SMART}\xspace}
\newcommand{\psherman}{{P-Sherman}\xspace}
\newcommand{\psmart}{{P-SMART}\xspace}

\usepackage{fix-cm}
\usepackage[colorinlistoftodos,prependcaption,textsize=small]{todonotes}
\usepackage{varwidth}
\def\thepapertitle{\name: Scalable Range Indexing on Disaggregated Memory [Extended Version]}

\def\thepaperkeywords{Disaggregated memory, B+-tree, offloading, pushdown, caching, indexing}

\hypersetup{
  pageanchor=true,
  plainpages=false,
  pdfpagelabels,
  bookmarks,
  bookmarksnumbered,
  pdfborder=0 0 0,  
  colorlinks=true,
  linkcolor=ACMRed,
  anchorcolor=ACMRed,
  citecolor=Blue, 
  pdftitle={\thepapertitle},
  pdfauthor={Baotong Lu, Kaisong Huang, Chieh-Jan Mike Liang, Tianzheng Wang, Eric Lo},
  pdfsubject={},
  pdfkeywords={\thepaperkeywords},
}

\AtBeginDocument{%
  }

\usepackage{tikz}
\newcommand*\circled[1]{\tikz[baseline=(char.base)]{
            \node[shape=circle,fill=.,inner sep=0pt] (char) {\color{white}\textsf\footnotesize #1};}}

\newcommand{\ib}{InfiniBand\xspace}
\newcommand{\rdmaread}{\textsf{\small RDMA READ}\xspace}
\newcommand{\rdmacas}{\textsf{\small RDMA CAS}\xspace}

\newcommand{\rdmawrite}{\textsf{\small RDMA WRITE}\xspace}
\definecolor{brickred}{rgb}{0.8, 0.25, 0.33}
\definecolor{BRICKRED}{rgb}{0.8, 0.25, 0.33}
\definecolor{BLUE}{rgb}{0, 0, 255}
\newcommand{\added}[1]{\textcolor{black}{#1}}
\newcommand{\move}[1]{\textcolor{black}{#1}}

\floatstyle{ruled}
\usepackage{algorithm}
\newfloat{algorithm}{t}{lop}

\newcommand\vldbdoi{XX.XX/XXX.XX}

\newcommand\vldbauthors{\authors}

\newcommand\vldbpagestyle{plain}

\begin{document}

\renewcommand\thefigure{\Alph{figure}}    
\renewcommand\thetable{\Alph{table}}    

\title{\thepapertitle}

\settopmatter{authorsperrow=3}
\author{Baotong Lu}
\authornote{Work partially performed while at The Chinese University of Hong Kong and Simon Fraser University.}
\affiliation{\institution{Microsoft Research}}
\email{baotonglu@microsoft.com}

\author{Kaisong Huang}
\affiliation{\institution{Simon Fraser University}}
\email{kha85@sfu.ca}

\author{Chieh-Jan Mike Liang}
\affiliation{\institution{Microsoft Research}}
\email{liang.mike@microsoft.com}

\author{Tianzheng Wang}
\affiliation{\institution{Simon Fraser University}}
\email{tzwang@sfu.ca}

\author{Eric Lo}
\affiliation{\institution{The Chinese University of Hong Kong}}
\email{ericlo@cse.cuhk.edu.hk}

\newcommand\repourl{https://github.com/baotonglu/dex}

\renewcommand{\shortauthors}{\vldbauthors}
\renewcommand{\vldbauthors}{Baotong Lu, Kaisong Huang, Chieh-Jan Mike Liang, Tianzheng Wang, Eric Lo}

\begin{abstract}
Memory disaggregation can potentially allow memory-optimized range indexes such as B+-trees to scale beyond one machine while attaining high hardware utilization and low cost.  
Designing scalable indexes on disaggregated memory, however, is challenging due to rudimentary caching, unprincipled offloading and excessive inconsistency among servers.

This paper proposes \name, a new scalable B+-tree for memory disaggregation. 
\name includes a set of techniques to reduce remote accesses,  
including logical partitioning, lightweight caching and cost-aware offloading.  
Our evaluation shows that \name can outperform the state-of-the-art by 1.7--56.3$\times$, and the advantage remains under various setups, such as cache size and skewness. 
\end{abstract}

\maketitle
\pagestyle{\vldbpagestyle}
\begingroup
\renewcommand\thefootnote{}\footnote{\noindent
This document is an extended version of "\name: Scalable Range Indexing on Disaggregated Memory", 
which will appear in The 50th International Conference on
Very Large Data Bases (VLDB 2024). This document is freely provided under Creative
Commons.
}\addtocounter{footnote}{-1}\endgroup


\renewcommand\thetable{\arabic{table}}    
\renewcommand\thefigure{\arabic{figure}}    
\setcounter{figure}{0}   
\setcounter{table}{0}   
\setcounter{section}{0}
\setcounter{page}{1}

\section{Introduction}
\label{sec:intro}

Memory-optimized indexes~\cite{ART,Masstree,HOT,AdaptiveHybridIndex,buzzword} are crucial for accelerating OLTP. 
Their scalability and economy, however, are being limited by the traditional monolithic server architecture where CPU and memory (DRAM) are ``bundled'' together. 
Blindly scaling up can lead to high cost that often only pays off under the full load. 
Worse, as data size---and consequently index size---grow, the demand for memory capacity can go beyond what a single server could offer. 
Memory disaggregation~\cite{zhang2020understanding,dmisca,memcompute,ThymesisFlow} has emerged to ease this problem by separating memory and compute into their own server pools and interconnecting the two resource pools via fast networks (e.g., \ib (IB)~\cite{IB} and CXL~\cite{CXL}). 
As the workload changes, we have the flexibility to independently scale compute threads and memory size, achieving high utilization and low cost.

\begin{figure}[t]
\centering
\includegraphics[width=\columnwidth]{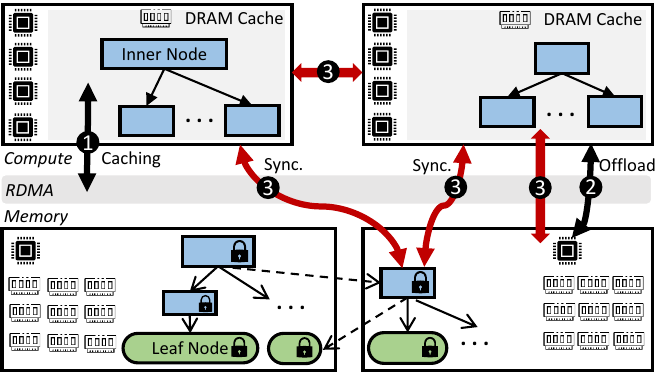}
\caption{\added{\textmd{Desiderata of indexes on disaggregated memory. 
\protect\circled{1} Caching should work with the smaller speed gap between local and remote memory, and limited local memory.
\protect\circled{2} Offloading should be aware of the scarcity of memory-side compute.
\protect\circled{3} Design should recognize potential data inconsistencies among servers (red arrows).}}
}
\label{fig:dm} 
\end{figure}

\subsection{Range Indexes: Disaggregated $\neq$ Scalable}
\label{subsec:index-dm}

\added{
Unfortunately, na\"ively deploying a tree index on disaggregated memory does not 
automatically achieve the aforementioned goals. With compute and memory decoupled, accessing
the index inherently incurs remote memory accesses (e.g., over RDMA), which can
add non-trivial latency (as compared to local DRAM accesses). This problem
exacerbates, as we consider that an index operation (e.g., lookup) typically
requires traversal from the root to the leaf node, necessitating at least one
RDMA operation per tree level.}

\added{
Memory disaggregation brings unique solution design space to the problem above.
Although compute and memory are decoupled, there is actually a hidden resource
hierarchy --- compute servers can have some local memory (in addition to the
larger remote memory), and memory servers can have some local compute (in
addition to the more powerful compute servers).} 

\added{
At first glance, leveraging local resources can reduce remote memory
accesses, through compute-side software-managed caching or offloading (aka computation pushdown) to memory servers.\footnote{We use offloading and computation pushdown interchangeably in this paper.} 
However, we argue that disaggregation mandates a significant departure from 
existing caching and offloading approaches~\cite{lruk,rdmaindex,smart,sherman}, 
to address the unique challenges in Figure~\ref{fig:dm}:
\protect\circled{1} rudimentary caching,
\protect\circled{2} unprincipled offloading,
and \protect\circled{3} excessive inconsistency.} 

\added{
\textbf{Rudimentary Caching.}
Given the availability of compute-side memory, it is natural to
cache frequently accessed index nodes on
compute servers.\footnote{Unless otherwise noted, 
throughout this paper ``cache'' refers to the software-controlled DRAM cache on compute servers, instead of CPU caches.}  
Many efforts have focused on improving cache-hit ratio~\cite{lruk,2q}. 
These past approaches, however, were mostly targeting conventional monolithic DBMSs.
Disaggregation invalidates certain assumptions of
traditional caching mechanisms.}

\added{
First, the speed gap between local and remote memory ($\sim$$10\times$ through RDMA) is much narrower than
that between memory and storage (e.g., $\sim$$1000\times$ with SSDs). Therefore, software
overheads associated with cache maintenance and synchronization become more
prominent in the disaggregation setting. 
}


\added{
Second, compute servers should not assume a certain amount of
local memory, especially the industry has not converged to one particular
disaggregation practice~\cite{ewais_survey}. As such, if local memory is
severely constrained, general-purpose caching mechanisms that do not exploit the
properties of tree indexes might not perform well.
}

\added{
\textbf{Unprincipled Offloading.}
%
Offloading takes advantage of the limited 
compute on memory servers for less RDMA communication. Conceptually, a compute 
thread simply sends \textit{one} index operation to a memory server, instead of 
individual RDMA operations that incur multiple round trips. However, the 
disaggregation setting now requires us be resource-aware to avoid overwhelming the 
limited compute on memory servers. 
In other words, informed decisions (i.e., what, when, and how much to 
offload) become crucial, to avoid overloading and guarantee scalable
performance.}

\added{
\textbf{Excessive Inconsistency.}
Memory disaggregation brings the challenge of handling different sources of 
data inconsistencies. Even without considering caching and offloading, 
we need synchronization (e.g., locks) to guard the index from being concurrently modified 
by compute threads. Unfortunately, existing RDMA-based locking for distributed 
synchronization is costly~\cite{rdmaguide}. If we implement 
caching on compute servers, it becomes necessary to ensure the coherence 
among all compute-side caches. This complexity grows with the number 
of compute servers. 
The problem exacerbates if we consider both compute-side caching and memory-side 
offloading. Memory servers now become another potential source of data changes. 
Prior to serving an offloaded operation, the memory server needs to ensure its view 
of the tree index is consistent with all compute-side caches. Such global 
consistency needs to be guaranteed throughout offloading, with the use of locks 
or coherence messages. Finally, all compute-side caches that contain stale pages 
should be synchronized with memory servers.}

\subsection{\name}
\label{subsec:dex}

This paper presents {\name}, a new B+-tree designed to scale on disaggregated memory.
\name uniquely combines a set of new and existing techniques to effectively mitigate the aforementioned issues.

\textbf{Compute-Side Logical Partitioning.}
\name mitigates cross-compute consistency overhead using logical partitioning~\cite{DORA,dinomo} where each compute server logically ``owns'' a set of key ranges while the memory servers still present a globally addressable shared space. 
This way, different compute servers operate mostly on disjoint portions of the index, reducing the cost of \added{cache coherence} across compute servers and \added{RDMA-based synchronization for remote 
memory accesses}. 
Load balancing and adding/removing a compute server are simple and lightweight since logical partitioning only necessitates adjusting routing without physically re-partitioning data.

\textbf{Optimized Caching.}
We propose a lightweight cache replacement strategy based on random sampling. 
\added{
By avoiding centralized data structures like FIFO queues, 
\name's cache reduces the contention and achieves high scalability. 
}
To reduce cache misses, we leverage application-level information to do
\textit{path-aware} caching that tends to keep in the cache frequently accessed \textit{index paths} from the root to lower level nodes. 
A child B+-tree node cannot be admitted to the compute-side cache unless its parent node has been cached. 
Similarly, in most cases, a parent B+-tree node is not evicted until all of its child nodes are evicted.
This not only improves cache efficiency (as nodes closer to the root are hotter) but also enables more effective offloading \added{with low consistency overhead between compute and memory servers (described below).}

\textbf{Opportunistic Offloading.}
{\name} tracks resource availability on memory servers at runtime, and it
offloads an index operation only if the completion time could be minimized.
\added{
However, a challenge arises from the simultaneous use of compute-side caching
above. Considering an index operation involving the tree traversal of
$N_{A}\rightarrow N_{B}\rightarrow N_{C}$, a na\"ive caching policy (e.g., the
widely used random eviction~\cite{watt, scalestore}) may evict $N_{A}$ 
and admit $N_{B}$ and $N_{C}$ into the compute-side cache. However, at this point, 
if there is another thread trying to offload the traversal of 
$N_{A}\rightarrow N_{B}\rightarrow N_{C}$ to the memory server 
after observing a cache miss on $N_{A}$,
concurrent changes made on the compute-side cache and the memory-side data 
could result in data inconsistencies.
Our design takes advantage of a property of
path-aware caching, where a consecutive path from the root to lower level nodes are cached. 
This effectively prevents a tree traversal from being
interleaved with caching and offloading, hence eliminating another source of potential
inconsistencies.}

The contributions of {\name} lie in systematically realizing an unique
combination of compute-side caching and memory-side offloading, in order to best
minimize the scalability bottleneck on disaggregated memory (i.e., remote memory
accesses). Evaluations on a four-server RDMA cluster show that {\name}
outperforms state-of-the-art, with 1.7--56.3$\times$ higher throughputs, under
various workloads. 

\section{Background and Motivation}
\label{sec:background}
We give the necessary background on disaggregated memory and its impact to tree indexes that motivated our work. 

\subsection{Disaggregated Memory (DM)}
\label{subsec:dm}
Compute (CPU cores) and memory (DRAM) have been traditionally coupled to scale together in data centers. 
However, as shown by recent work~\cite{alibaba,borg}, this can lower memory utilization and waste compute resources. 
As networking technologies advance, it now becomes viable to decouple compute and memory respectively into compute and memory servers that can independently scale, similar to how storage is disaggregated in the cloud. 

A typical disaggregated memory architecture consists of a set of compute servers and a set of memory servers, as Figure~\ref{fig:dm} shows (ignore the tree nodes for now). 
Compute servers focus on providing ample compute capabilities with high core count and high CPU frequency; their memory capacity is usually limited. 
Memory servers focus on providing ample memory capacity, but their compute capabilities are limited.   
A high-speed interconnect such as \ib and CXL allows compute servers to access data on memory servers using memory semantics. 
Currently, this is widely done using RDMA over \ib, although 
other solutions (e.g., CXL.mem~\cite{CXL}) are being devised. 
We follow recent work~\cite{rdmaindex,sherman,smart,zhang2020rethinking,zhang2020understanding,dinomo,rolex} to focus on RDMA-enabled disaggregated memory. 

RDMA allows participating servers to access each other's memory directly without involving the remote CPU and/or OS kernel, providing much lower latency than TCP/IP networks. 
RDMA is performed by ``verbs'' which can be one-sided (\texttt{READ}, \texttt{WRITE} and atomics such as compare-and-swap or \texttt{CAS}) or two-sided (\texttt{SEND}/\texttt{RECV}).
One-sided verbs do not involve remote CPU,
whereas two-sided verbs operate similarly to TCP/IP operations by requiring the remote CPU to participate.
Due to limited memory-side compute power, one-sided verbs are preferable for many disaggregated memory settings.
\added{
    Nonetheless,
    offloading can be achieved by performing remote procedure calls (RPCs) using either one-sided~\cite{farm} 
    or two-sided verbs~\cite{FaSST, rdmaindex}. 
    Either approach requires memory-side threads to receive and process RPC requests.
    Following recent research~\cite{rdmaindex, FaSST}, we use two-sided verbs for offloading, as it achieves better performance.
}

\subsection{Disaggregating Memory-Optimized Indexes} 
\label{subsec:tree-on-dm}
With the aforementioned architecture, now we discuss how software, in particular tree-based range indexes, can be adapted. 
Without losing generality, we focus on memory-optimized B+-trees that are designed for multicores assuming the tree fits in memory, and use memory-optimized layout and optimistic locking~\cite{olc,optiql} or lock-free concurrency~\cite{BwTree}. 
Importantly, most of them are shared-everything where any thread can access any part of the tree, which recent DM-based B+-trees have inherited.  
As shown in Figure~\ref{fig:dm}, these properties allow (1) using remote memory as ``the main memory'' to store tree nodes and (2) using the CPU cores in compute servers to perform tree operations. 

To probe for a key, the compute server issues a one-sided \rdmaread to fetch the root node to its local DRAM.  
It then searches the node to find the next child node, which again is fetched to the compute server's local DRAM using \rdmaread. 
Depending on how the tree nodes are distributed across memory servers, a traversal may involve multiple memory servers.  
To coordinate accesses to shared data on a memory server, optimistic locks are replaced with RDMA-based locks built using atomics such as \rdmacas.
Also, since there is a non-trivial speed gap between accessing local DRAM and remote memory, it becomes important to cache frequently/recently accessed nodes in the compute server. 

RDMA brings two major challenges. 
(1) RDMA does not provide off-the-shelf coherence among servers, leaving the responsibility of ensuring data consistency including data synchronization and cache coherence across servers to the implementation.
As a result, after a compute server updates a node using \rdmawrite, the cached node in other compute servers become stale and should be invalidated, which is typically done by explicitly sending coherence messages across compute servers. 
(2) RDMA exhibits higher latency ($\sim$2000ns) than local DRAM ($\sim$100ns). 
Both challenges require disaggregated indexes to avoid unnecessary remote accesses, discussed next. 

\subsection{State-of-the-Art and Motivation}
\label{sec:sota}

Recent work goes beyond the naive adaptation to reduce RDMA operations. 
Section~\ref{subsec:index-dm} has covered some of the issues, here we analyze in detail the design of two representative designs---\sherman~\cite{sherman} (a DM-optimized B+-tree) and \smart~\cite{smart} (a DM-based trie~\cite{ART})---and how they still do not scale well, which motivated our work.

\textbf{Compute-Side Caching.}
A shared-everything disaggregated index normally needs to implement cache coherence by exchanging coherence messages across compute servers (e.g., to invalidate stale nodes). 
Both \sherman and \smart observed that exchanging coherence messages is costly as it can be as expensive as cache misses. 
To avoid such cost, they do not cache leaf nodes and only cache inner nodes, for which coherence is not strictly required: 
using stale cached inner nodes during a traversal will not read inconsistent data but lead to incorrect leaf nodes, which can be easily resolved by retrying the operation and fetching the up-to-date index nodes from the memory pool. 
\sherman only caches the lowest levels of inner nodes 
and builds an extra index for cached nodes in compute servers. 
As a result, 
it always incur one RDMA operation to access a leaf node, necessitating RDMA even when the cache has enough capacity to hold all the needed nodes. 
Maintaining the extra index also requires extra bookkeeping. 
Similar observations apply to \smart.
\added{
In other words, existing work trades the benefits of caching leaf nodes
for reducing coherence overheads. 
}


In terms of caching, prior work~\cite{leanstore,2q,lruk} primarily focused on buffer pool solutions targeting the DRAM-SSD/HDD hierarchy.
The large gap between DRAM and HDDs/SSDs means that data movement cost (storage I/O) is the major bottleneck, justifying the use of simple centralized data structures to maintain cache metadata (e.g., centralized LRU lists). 
However, in disaggregated memory, the latency gap between compute-side DRAM and remote memory is relatively small. 
Moreover, compute servers in the DM-setting should have even higher core counts, putting significant pressure on the caching data structure and replacement mechanisms. 

\added{
We observe that, with lower latency gap between local and remote memory, 
cache replacement frequency becomes a critical factor of the synchronization cost.
In our analysis,
we denote the latency to access a cached and uncached page
as $T_{c}$ and $T_{d}$, respectively.
For simplicity, we do not consider bandwidth limits.
Suppose the cache miss ratio is $R$,
the frequency can be computed as: 
$Replacements/second=\frac{1~second}{T_{d} + (\frac{1-R}{R}) \times T_{c}} \times N_t$, 
where $N_t$ is the number of threads. 
We empirically set $T_{c}$ as \SI{400}{\nano\second} 
for the access to a 1KB DRAM-resident cached page. 
$T_{d}$ for SSDs is typically 100$\mu$s~\cite{ssdlatency} while it takes 
2$\mu$s for one \rdmaread. 
At the same cache miss ratio (e.g., 10$\%$) and number of threads (e.g., 36),
the replacement frequency for DRAM-SSD is $0.35 \times 10^6 $ while it is $6.43 \times 10^6$ 
for disaggregated memory, such that there is over $18 \times$ higher. 
Even worse, given the limited memory capability in compute servers, 
we can easily get higher $R$ and thus higher replacement frequency.  
}

Existing DM-based indexes~\cite{sherman,smart} did not take these into consideration. 
For example, upon cache admission, \smart needs to update a centralized local counter to track cache usage and uses a centralized FIFO queue to organize cache entries.  
\added{With high replacement frequency by concurrent threads,}
our evaluation in Section~\ref{sec:eval} shows that it exhibits severe contention and cannot scale.

\textbf{Offloading.} 
Although attractive, 
neither \sherman nor \smart considered offloading. 
Other existing approaches~\cite{rdmaindex} can offload full or partial index operations, but do so by hard-coding policies. 
Without considering the actual capabilities and load of memory servers, 
one may easily overload the memory servers, defeating the purpose of offloading. 
It is crucial to selectively offload index operations and strike a balance between remote accesses and offloading. 

\textbf{Consistency.} 
\added{Besides cache coherence,} 
to ensure correct \added{concurrent accesses to the memory pool}, \sherman and \smart use distributed optimistic locks~\cite{rdmaguide} \added{for synchronization}. 
These locks are the same as their monolithic counterparts, except the low-level implementations are based on RDMA primitives.
A write operation must obtain the lock in exclusive mode by atomically changing the lock word to the ``locked'' state using \rdmacas, while a read operation only needs to verify that the protected node did not change after the read operation. 
While the adaptation is simple, it turns out that the RDMA-based verification process is very heavyweight by incurring two \rdmaread operations, significantly lowering performance. 
Existing work (including \sherman) has overlooked this issue, leading to unsafe implementations~\cite{rdmaguide}. 
Moreover, DM-based indexes are more prone to performance collapse under skewed workloads~\cite{sherman} because RDMA-based locks incur high overheads due to multiple network roundtrips caused by retries.
Thus, designing scalable synchronization remains an open challenge. 

\added{
    Although no existing DM-based indexes support both caching and offloading, 
    some general frameworks~\cite{teleport} enable offloading 
    through system calls. 
    However, targeting fast memory-resident indexes, we prefer user-space solutions with low overhead.
}

\section{\name Overview}
\label{sec:design}
\name is a B+-tree optimized for disaggregated memory that combines a set of new and existing techniques in 
logical partitioning, compute-side caching and opportunistic offloading to mitigate the issues identified in previous sections. 

\begin{figure}[t]
\centering
\includegraphics[width=0.9\columnwidth]{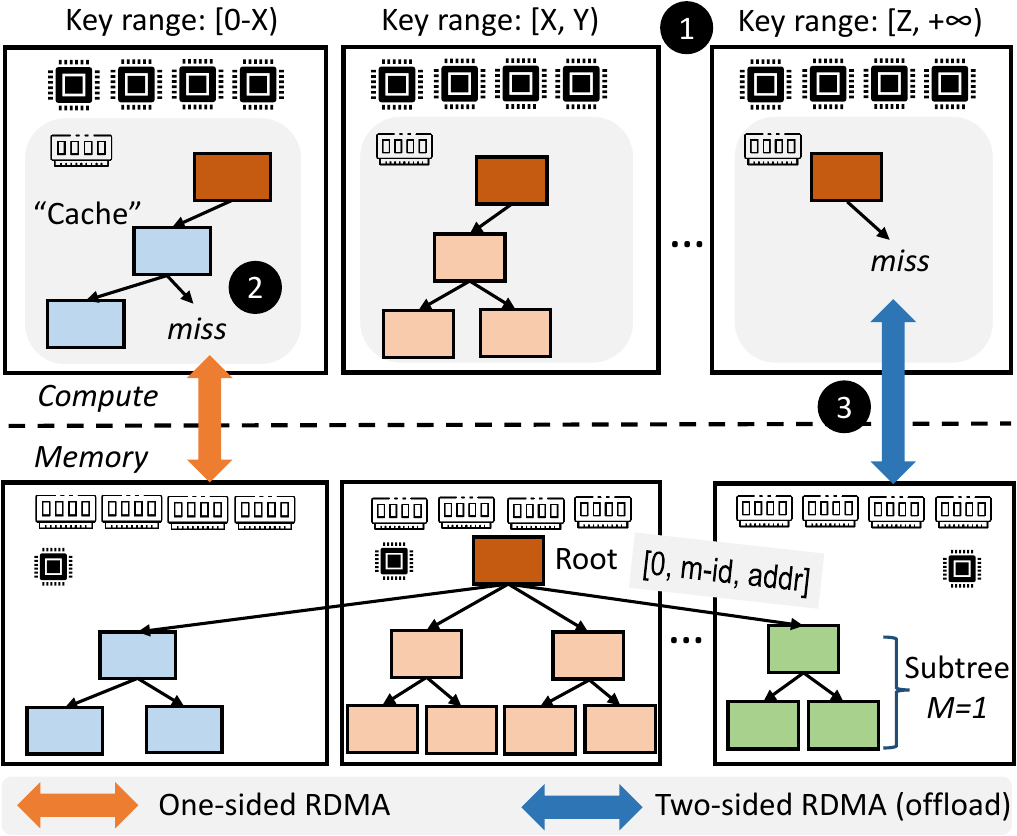}
\caption{\textmd{Overview of \name.  
\protect\circled{1} Each compute server ``owns'' a disjoint range of the key space and
\protect\circled{2} caches tree traversal paths in local DRAM.
\protect\circled{3} Upon cache misses, the compute server selectively offloads index operations when profitable. 
B+-tree nodes are distributed onto memory servers. However, subtrees under level $M$ are all located in the same memory servers to avoid expensive pointer chasing across memory servers during offloading.}
}
\label{fig:arch} 
\end{figure}

\textbf{Index Placement.} 
As a B+-tree variant, \name uses normal B+-tree nodes which are distributed onto different memory servers, 
as shown by Figure~\ref{fig:arch} (bottom). 
In particular, we group index nodes into sub-trees and ensure a subtree rooted at level $M$ ($M=0$ for the leaf level) is entirely stored on the same memory server.
For example, as shown in Figure~\ref{fig:arch}, all nodes in the subtrees rooted at level $M=1$ are all stored in the same memory servers. 
As we describe later, this facilitates better offloading. 

\textbf{Node Layout and Addressing.}
Each tree node begins with a header area (including metadata such as the lock), followed by a key array and a pointer array (for inner nodes) or a value array (for leaf nodes). 
Different from monolithic B+-trees,
the index nodes in \name need to form a unified, global memory address space among multiple memory servers, 
such that compute servers can locate index nodes in memory servers.   
Similar to previous work~\cite{sherman,smart,rdmaindex}, we address it using pointer tagging.  
A pointer is still 64-bit but is tagged with extra information in the format of \textsf{[swizzled, memory-server-id, address]}, leveraging the fact that modern 64-bit x86 microarchitectures do not implement all the 64 bits~\cite{IntelManual}.
The trailing 48-bit \texttt{address} carries a local memory address of a particular server identified by the 15-bit \texttt{memory-server-id}. 
The combination of \texttt{memory-server-id} and \texttt{address} form a unique global address in the memory pool. 
They are also stored in the header of each tree node to identify the node. 
The most significant \texttt{swizzled} bit indicates whether \texttt{address} is local to the current compute server, which is only used by the compute-side cache. 

\textbf{Index Node Accesses.}
With B+-tree nodes distributed across memory servers, each compute server may access tree nodes
via one-sided RDMA and possibly cache them on the compute-side. 
Different from prior solutions, \name is able to cache both inner and leaf nodes to better use the cache space. 
\name departs from shared-everything architectures and adopts logical partitioning on the compute side to sidestep the vast majority of cross-server coherence issues. 
Synchronization is also mostly only needed within each compute server without involving distributed locks, except for certain index nodes crossing logical partition boundaries (e.g., the root node in Figure~\ref{fig:arch}) that can be accessed by more than one compute server. 
As shown in Figure~\ref{fig:arch}(top), regardless of how the index nodes are distributed across the memory servers, each compute server logically ``owns'' a disjoint range of keys and is responsible for handling all the requests in that range. 


Upon a cache miss, the compute server will either fetch the missing node from or offload the remaining index operation to memory servers, depending on whether offloading is profitable. 
In the former case, \name combines a scalable cache replacement mechanism, path-aware caching and selective cache admission to efficiently cache tree nodes. 
As mentioned earlier, index nodes are grouped into sub-trees. 
Therefore, offloading the operation of traversing a subtree rooted at level $M$ will not cause remote pointer chasing across memory servers, improving offloading performance and reducing implementation complexity in memory servers. 

Next, we elaborate how \name realizes the above functionality by 
logical partitioning (Section~\ref{sec:partition}), compute-side caching (Section~\ref{sec:cache}), and opportunistic offloading (Section~\ref{sec:pushdown}). 

\section{Compute-Side Logical Partitioning}
\label{sec:partition}

\added{
As discussed in Section~\ref{sec:sota}, 
prior works~\cite{sherman, smart} made various trade-offs to mitigate cache coherence overhead, 
yet they still exhibit excessive remote accesses to leaf nodes and require costly RDMA-based synchronization. Therefore,}
\name uses logical partitioning to greatly reduce cross-compute consistency overhead, and to improve cache locality.  
As Figure~\ref{fig:arch} shows, each compute server owns a logical partition of keys.
\name can work with various range partitioning schemes (e.g., equal-width or workload-aware), as long as each leaf node is exclusively owned by exactly one partition (i.e., one compute server). 
This effectively limits the need for cross-server synchronization and cache coherence to only inner nodes, which are less frequently updated than leaf nodes. 
It is also easily achievable by picking partition boundaries using keys from the lowest inner node level because keys in these nodes indicate the possible key range of a leaf node 
(fence keys~\cite{fence}). 

\begin{algorithm}[t]
\begin{lstlisting}[language=python,
	mathescape,
	gobble=0,
	keywordstyle=\ttfamily\bfseries\color{blue},
	]
def cache::remote_read(node_addr, shared):
  if shared is true:
    version = RDMA_read(node_addr, 8B)
    if is_lock_set(version): return NULL
    node = RDMA_read(node_addr, node_size)
    if version != RDMA_read(node_addr, 8B): return NULL
  else 
    node = RDMA_read(node_addr, node_size)
  cached_node = insert_to_cache(node)
  return cached_node

def lookup(key):
retry:
  parent = NULL, vp = 0
  cur_node = cache.lookup(root)
  if cur_node is NULL:
    cur_node = cache.remote_read(root)
    if cur_node is NULL: goto retry
  vc = cur_node.version_lock
  if is_lock_set(vc): goto retry
  
  while(cur_node is inner):
    if parent != NULL and vp != parent.version_lock:
      goto retry
    parent = cur_node, vp = vc
    child_addr = parent.search(key)
    cur_node = cache.lookup(child_addr)
    if cur_node is NULL:
      shared = is_child_shared(child_addr, parent)
      # Consider operation offloading
      if shared is false and deserve_offload() is true:
          result = offload(child_addr, key)
          return result
      # Conduct caching
      cur_node = cache.remote_read(child_addr, shared)
      if cur_node is NULL: goto retry
    vc = cur_node.version_lock
    if is_lock_set(vc): goto retry
    if cur_node.fence_keys is not valid: 
      refresh_from_root() and goto retry
  
  result = cur_node.lookup(key)
  if parent != NULL and vp != parent.version_lock:
    goto retry
  if vc != cur_node.version_lock: goto retry
  return result
\end{lstlisting}
\caption{\added{\name lookup algorithm.}} \label{algo:lookup}
\end{algorithm}

\added{
Since certain nodes like the root are shared (i.e., crossing partition boundaries) 
by multiple compute servers while the others are not, we handle the concurrency control of them in different ways.
Within a compute server, accesses to cached tree nodes are synchronized using in-memory optimistic lock coupling~\cite{olc}, 
as shown in Algorithm~\ref{algo:lookup} \added{(lines 19--20, 23--24, 37--38, 43--45).}
Upon a cache miss, \name either offloads the remaining index operation to the memory server (line 32) 
or retrieves the missing node from the memory pool to the compute side (line 35) through the \texttt{cache::remote\_read} function.
    The \texttt{cache::remote\_read} function determines whether to perform cross-server synchronization 
    based on whether the target node crosses partition boundaries. Specifically, if the target node is shared by multiple compute servers, 
    \name utilizes RDMA-based optimistic locking~\cite{rdmaguide} to synchronize concurrent accesses (lines 3--6). 
    Otherwise, it simply reads the target node through a single \rdmaread operation (line 8).
}

Nonetheless, such cross-server synchronization is infrequent 
because only few inner nodes would cross the partition boundaries.
Moreover, inner nodes that span across partition boundaries are typically closer to the root and are therefore more likely to be cached on the compute-side,
largely eliminating the need for reading them with RDMA-based synchronization.
However, for the updates to shared inner nodes, 
RDMA-based locking is still required and
\name writes back their updates to the memory pool to ensure the memory pool always has the most up-to-date version.

Cross-compute cache coherence is only required for inner nodes that cross a logical partition boundary \textit{and} are cached. 
Since inner nodes store no data but guiding information, \name follows prior work~\cite{sherman} to only bring in the fresh copy of an inner node from remote memory when its staleness lands its search to a wrong child node.
To detect such cases, \name records fence keys of the index nodes into their headers. 
Upon accessing a child node, \name checks if the search key is within the fence keys in the header.
If not,
\added{as shown in lines 39--40 of Algorithm~\ref{algo:lookup},}
\name will restart the search by bringing in fresh nodes on the search path from the remote root and invalidating stale cached nodes. 
\added{
Note that logical partitioning does not change the worst-case complexity of remote memory accesses. 
For instance,
in an extreme scenarios where local resources (e.g., cache) are minimal or when refreshing cache from the remote root, 
\name still necessitates $O(h)$ remote accesses, where $h$ represents the height of the tree.  
}

\added{
Although logical partitioning helps reduce cross-compute synchronization and cache coherence overhead, 
it can incur load imbalance in which certain compute servers are overloaded. 
However, this can be greatly mitigated through logical re-partitioning. 
}
Different from physical re-partitioning which requires expensive data movement and index rebuild,
logical partitioning enables \name to re-partition the overloaded compute server by simply 
re-adjusting the boundaries of logical partitions, without any physical data movement.  
Upon re-partitioning, the involved compute servers simply flush their dirty cache pages to the memory pool and adjust the partitioning boundaries.
Experiments in Section~\ref{sec:eval} show that re-partitioning in \name is lightweight and can be finished within seconds.  
We leave load balancing policy questions~\cite{autoscale}, such as when compute servers should be scaled, 
as interesting future work.  
\added{
Another advantage is that elasticity (scale-in and scale-out) can be supported very easily 
through lightweight logical re-partitioning. 
}
\section{Compute-Side Caching}
\label{sec:cache}
\name deploys a cache for B+-tree nodes (both inner and leaf) on each compute server using its local DRAM. 
Thanks to logical partitioning, communication between compute-side caches is rare, allowing us to localize the problem of optimizing caching on a single compute server.
Since recent DBMS buffer pools optimized for fast SSDs share similar scalability concerns~\cite{leanstore,lean-evolution}, we design \name cache based on these techniques to attain a reasonable baseline, on top of which we propose optimizations specific for disaggregated B+-trees. 

\begin{figure}[t]
    \centering
    \includegraphics[width=0.9\columnwidth]{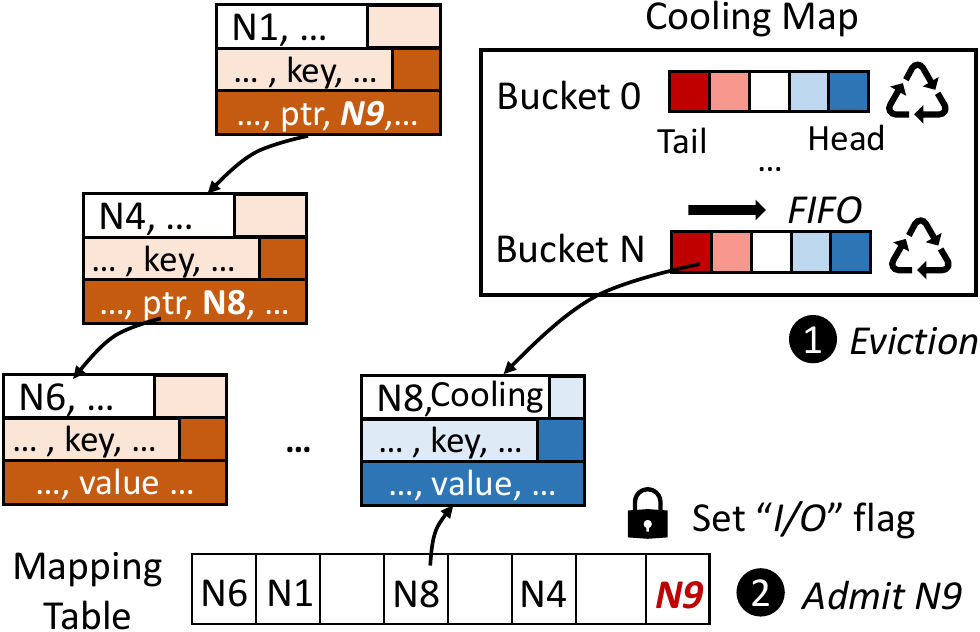}
    \caption{\textmd{\name caching in a compute server.
    \protect\circled{1} Potential eviction candidates are first admitted to the cooling map which is a hash table of FIFO arrays to alleviate contention. 
\protect\circled{2} To admit a new node (N9), the first thread that accesses it signals in-progress RDMA by atomically setting an \texttt{I/O} flag in the mapping table. 
Subsequent concurrent threads will then re-traverse the path from root to avoid repeatedly issuing RDMA by multiple threads for the same node. 
    }}
    \label{fig:buffer} 
\end{figure}


\subsection{Overall Structure}
\label{sec:cache-struct}
Like its monolithic counterparts, as Figure \ref{fig:buffer} shows, \name tracks cached B+-tree nodes using a mapping table implemented using a concurrent hash table that maps node IDs (i.e., the global address of this node) to local node addresses in the cache. 
Each page frame in the cache is set to exactly the size of a B+-tree node, plus a header that embeds an optimistic lock to coordinate concurrent accesses from the same compute server.  
To access a node, a worker thread checks if it has been cached by probing the mapping table. 
Further, we follow representative buffer pool designs~\cite{leanstore} to introduce a cooling status for nodes that are identified as potential eviction candidates. 
These nodes are indexed by a cooling map that is looked up when eviction is necessary. 

To reduce the overhead caused by probing and modifying the mapping table, \name uses pointer swizzling~\cite{bigdata,leanstore}. 
Upon the cache admission of a child node $C$, we record in its parent node $P$ (which is already in the cache) $C$'s local address with \texttt{swizzled} bit set. 
This way, subsequent traversals of the same (sub)path will primarily involve only local pointer chasing. 

To reduce the overhead of selecting eviction candidates, \name uses a randomized cooling policy~\cite{leanstore} that differentiates hot and cold nodes in a coarse-grained manner.
When a compute thread finds its thread-local free-page set is empty, it randomly samples a set of nodes in the cache (two in our implementation), unswizzles them from their parent nodes, writes back the dirty pages to the memory pool, and sets their state to \texttt{cooling}, indicating that the page is a candidate for future eviction.

If the target node is \texttt{cooling} (i.e., was chosen as a candidate for future eviction), 
we restore its status to \texttt{cached} and re-swizzle its pointer in the parent node. 
This gives a second chance for frequently accessed pages that are accidentally sampled for cooling to stay in the cache.
In contrast, ``truly'' cold nodes will likely remain in the cooling state and are eventually evicted from the cache, improving cache hit ratio and reducing remote memory accesses. 

With the overall structure laid out, 
in the rest of this section, we propose further optimizations specific to disaggregated B+-trees.

\subsection{Scalable Cache Replacement}
\label{sec:replace}
\begin{figure}[t]
    \centering
    \includegraphics[width=0.9\columnwidth]{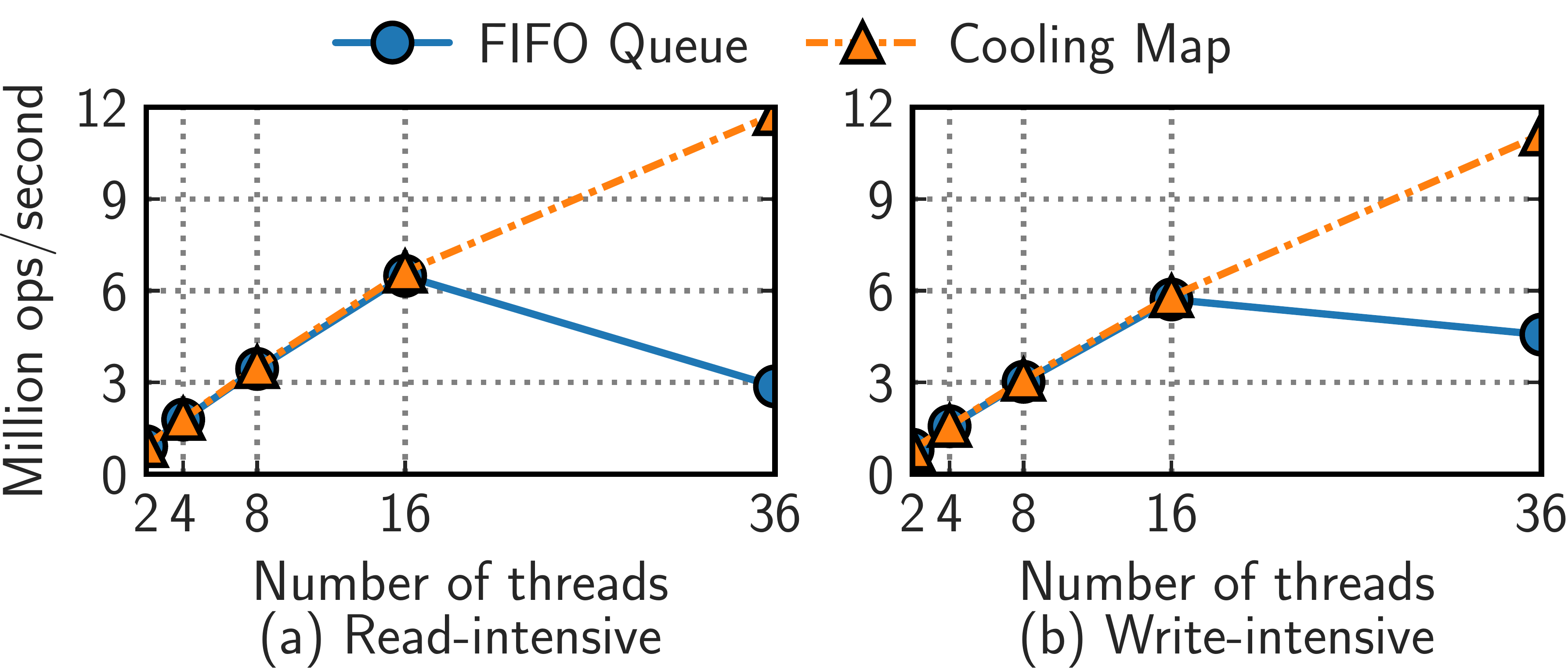}
    \caption{\textmd{\added{\name's scalability with different cooling structures.}}}
    \label{fig:cooling-map-evaluation} 
\end{figure}
\added{
The analysis in Section~\ref{sec:sota} shows that
the disaggregated memory setting can lead to much higher cache replacement frequency than DRAM-SSD settings. }
This in turn puts much more pressure on the data structure that tracks eviction candidates in the cache. 
Specifically, 
a fair eviction requires ranking hotness among cooling nodes.
Yet prior approaches usually use a centralized shared FIFO list~\cite{leanstore}
\added{protected by a single lock} to achieve this goal.
\added{
With high replacement frequency, 
threads need to frequently sample cooling pages to the shared FIFO queue 
or evict pages from it. 
This leads to intensive updates of the head/tail pointers of the queue, incurring
severe cache-line pingpong between CPU cores. 
}
\added{
As Figure~\ref{fig:cooling-map-evaluation} (details in Section~\ref{sec:setup}) shows, 
\name with FIFO queue cannot scale due to high synchronization cost 
in the cooling process. 
}

\name solves this problem with an extremely simple but effective tweak that replaces the FIFO list with a concurrent hash table where each bucket includes a FIFO array (thus called ``cooling map''). 
Because of the randomized nature of hash tables, this allows amortizing accesses over multiple memory locations, greatly alleviating contention.
As shown in Figure~\ref{fig:buffer}, each bucket includes a CPU cacheline-sized FIFO array, where each slot stores the local pointer to a cooling node. 
Each bucket is protected by a lock for correct multi-threaded accesses. 
When a cached node is selected for cooling, it is hashed into one of the buckets using its node ID and inserted into the tail slot of the FIFO array by shifting existing entries. 
If the array is already full, shifting would cause the head page to be evicted from the cache.
The evicted page is then inserted into a thread-local free page set. 
Therefore, with the cooling map, cache evictions in different buckets execute independently, significantly improving scalability. 
\added{
    The cooling map design strikes a balance between maintaining FIFO ordering (within each bucket) and scalability.
As shown in Figure~\ref{fig:cooling-map-evaluation}, \name with the cooling map scales well and outperforms the queue-based design.
}
Finally, as stated in Section~\ref{sec:cache-struct}, \name proactively unswizzles the pointer to the node once it is selected for cooling (e.g., N8 in Figure \ref{fig:buffer}). 
This allows worker threads to avoid the unswizzling process for the eviction page, simplifying the eviction 
process.

\subsection{Path-Aware Caching}
\label{subsec:path-aware-cache}
\name employs a path-aware strategy for cooling inner nodes where a tree path in a non-cooling state is always cached consecutively from the root node to low-level tree nodes.
Specifically, when selected for cooling (e.g., N4 in Figure \ref{fig:buffer}), an inner node will attempt to delegate the cooling command to one of its swizzled children (e.g., N6) instead. 
The cooling command will be recursively delegated until reaching 
a node with no swizzled child pointer (hence the leaf nodes are the base case).  
This way, we can avoid writing back nodes carrying swizzled pointers that are invalid in remote memory servers.
More importantly, such path-aware caching ensures that only the node at the end of the cache path 
will be transferred to a cooling state (i.e., becoming an eviction candidate), even if 
a node in the middle of the cache path was initially sampled.
While delegation is not new~\cite{leanstore,nvmbuffer}, 
as we discuss later in Section~\ref{sec:pushdown}, delegation in \name also enables more efficient pushing down of the remaining operation to the memory pool.

\subsection{(Selective) Cache Admission}
When a cache miss occurs, the compute thread needs to fetch the missing node from the memory pool and admit it into the cache.
The first thread that accesses the missing node will insert the node ID into the mapping table with an \texttt{I/O} flag as the value.
Subsequent threads that see \texttt{I/O} in the mapping table entry will re-traverse from the locally cached root to avoid repeatedly trying to admit the same node from remote memory. 
For example, as shown in Figure~\ref{fig:buffer}, admitting the child node $N9$ of the root node $N1$ 
requires setting its \texttt{I/O} flag in the mapping table before cache admission.  
This ensures that other concurrent threads that are also trying to load $N9$ will not attempt to issue more RDMA operations, saving network bandwidth. 
To admit a new node, the compute thread also needs to obtain a free page in its local free page set. 
If there are no more available local free pages, 
the thread will randomly select a bucket in the cooling map, evict the oldest page in the array, and use it as the free page for accommodating the incoming node. 
Thanks to the aforementioned fine-grained locking in the cooling map, multiple threads can obtain free pages from the cooling map without scalability bottlenecks.

Given that remote accesses are expensive and cache sizes are limited, it is important to keep hot nodes in the cache as long as possible. 
Yet previous disaggregated indexes~\cite{sherman,smart} employ an eager admission policy that unconditionally admits all the accessed nodes to the cache. 
This can result in the eviction of hot nodes in favor of newly retrieved nodes whose hotness has not yet been confirmed. 
Moreover, admitting newly retrieved nodes eagerly may introduce unnecessary overhead since provisioning free cache pages may trigger RDMA writeback for page eviction.

\name proposes a lazy admission policy to mitigate this issue. 
We assign each newly retrieved node a probability ($P_A$) of being admitted into the cache.  
Pages that cannot be admitted into the cache are discarded (or written back if dirty) immediately after use.
Based on experimental results, we empirically set $P_A$ to 0.1 for leaf nodes but set $P_A$ to 1 for inner nodes. 
In other words, inner nodes are still always admitted into the cache.
We made this design choice because 
inner nodes are generally hotter than leaf nodes. 
Besides, since \name adopts the lightweight optimistic locking for tree traversal, missing some nodes on a cached path 
would cause unnecessary complexity and overhead.
Exploring auto-tuning/learned techniques \cite{zhang19tuning,li19tuning} to determine $P_A$ is future work.

\section{Opportunistic Offloading}
\label{sec:pushdown}

\name improves resource utilization and reduces unnecessary remote accesses by offloading selected index operations to memory-side CPUs.  
This is enabled by \name's path-aware caching which exhibits an important property: 
upon a cache miss during a traversal at level $L$ of the tree, it is likely that subsequent node traversals from level $L-1$ down to the leaf (level 0) will also encounter cache misses. 
This is because \name employs cooling delegation (Section~\ref{subsec:path-aware-cache}) where a parent becomes an eviction candidate (i.e., in cooling) after all its children.
Therefore, \name makes the offloading decision upon a cache miss on node $N$ at level $L$, then the memory-side thread will take over and finish the remaining index operation including the local traversal from level $L$ to level $0$ and return the operation result to the compute server. 
However, \name does not offload (1) the traversal of inner nodes that do not exclusively belong to one partition or (2) when the node is at level $L > M$. 
Condition (1) ensures that the missed subpath from $N$ to the leaf level is dedicated to one compute server. 
For condition (2),
recall that \name enforces nodes in the sub-tree rooted at level $M$ are all stored in one memory server (Figure \ref{fig:arch}). 
These two conditions guarantee that offloading is confined between one compute server and one memory server, avoiding much complexity and remote pointer chasing across memory servers. 
For the same reason, \name will fall back to the normal path when an offloading attempt returns and reports that it would trigger a \added{structural modification operation (SMO)} on the memory server because SMOs (such as leaf node splits) can propagate up to nodes at level $M+1$ or higher.

\begin{figure}[t]
  \centering
  \includegraphics[width=0.9\columnwidth]{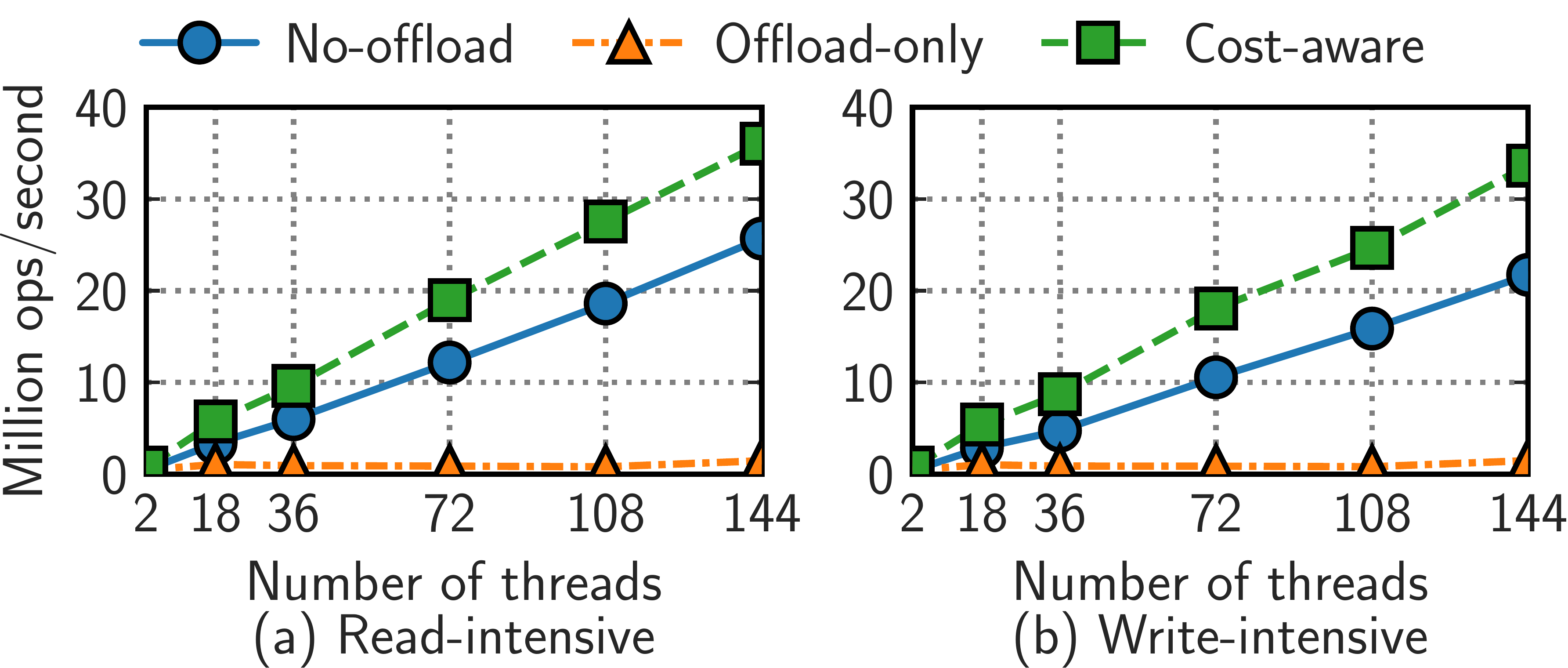}
  \caption{\added{\textmd{\name's throughput 
  under different offloading policies, 
  the cache size is set to $1\%$ of the data.}}}
  \label{fig:offload-evaluation} 
\end{figure}

\added{
It is important to note that offloading is only beneficial when there is spare compute capacity available in the memory pool.
Previous offloading policies~\cite{rdmaindex} that always offload index operations, can easily saturate limited memory-side CPU. 
We show this effect using an {\sf Offload-only} variant that caches inner nodes above level $M$ and always offloads the remaining index operation.
As shown in Figure~\ref{fig:offload-evaluation} (details in Section~\ref{sec:setup}), 
{\sf Offload-only} cannot scale due to high contention in the memory server caused by excessive requests, 
whereas our cost-aware approach (described below) can truly benefit from offloading 
and outperform other baselines.}

Next, we discuss how \name makes the offloading decision upon a cache miss, followed by our approach to ensuring consistency between compute-side cache and nodes in memory servers that are simultaneously modified by offloaded operations.

\subsection{Load and Cost Aware Offloading} 
\label{sec:load-aware}
\name determines whether a sub-path traversal is worth being pushed to remote memory using statistics collected at runtime. 
While more sophisticated methods (e.g., those that leverage machine learning) exist, \name strikes a balance between runtime overhead and decision-making accuracy.

Upon a cache miss on a node $N$ at level $L$, \name compares the estimated latencies of conducting the access via one-sided RDMA vs. offloading. 
The former is estimated as $(L + 1)\times (l_o+ l_s)\times c$, where $l_{o}$ is the latency of an \rdmaread, $l_s$ is the latency of a local node search in the compute-side cache, and $c$ is an empirically determined coefficient ($>$1) that accounts for the operational cost of using the compute-side cache 
(e.g., the cost of free page provisioning).
Our experiments show that $l_{s}$ and $c$ are relatively stable and are therefore specified upon tree initialization. 
The latter (offloading latency, $l_{p}$) and $l_{o}$ are empirically determined based on the moving average of a predefined number (e.g., 50) of recent samples of the corresponding actions (i.e., two-sided RPC for offloading and \rdmaread). 
To cope with workload changes, \name ensures that it has a small probability $q$ (e.g., 1\%) of taking the contrary action, allowing for regular updates.  
With these, an offloading request is issued only when offloading takes shorter latency than the accesses via \rdmaread, i.e., when $l_p< (L + 1)\times (l_o+ l_s)\times c$.

\subsection{Compute-Memory Data Coherence}
With concurrent updates from compute-side worker threads and memory-side offloading threads, tree nodes in the cache and memory pool may diverge, jeopardizing correctness. 
Therefore, data coherence needs to be resolved vertically~\cite{teleport} between compute and memory servers. 
\name first simplifies the problem by ensuring that offloading occurs exclusively between a pair of compute and memory servers, as stated earlier in this section. 
We then focus on ensuring coherence between one compute and one memory server.

A pushdown thread in a memory server by design always can operate on the latest subtree $T_N$ whose root is $N$, with no dirty pages of $T_N$ in the compute server's cache.
That is because when $N$ is not in the cache, it implies all nodes in $T_N$ are either evicted or under cooling (with their dirty pages written back to the memory pool already).
This mandates two conditions for correct offloading. 
(1) No concurrent compute threads are working on the cached (cooling) nodes in $T_N$ or using RDMA to retrieve any missing node in $T_N$ on the compute side when offloading is in-progress. 
(2) Any node update during offloading is propagated back to the compute node to invalidate the corresponding cached copies. 

To satisfy condition 1, before the offloading operation starts, \name pins $N$'s parent in a hot state to safeguard it against eviction. 
In addition, we insert $N$ into the mapping table with the \texttt{I/O} flag as the value. 
Concurrent compute threads seeing the \texttt{I/O} flag when looking up $N$ in the mapping table would restart from the root to avoid conflicts with the concurrent ongoing pushdown request, similar to how we prevent concurrent threads from issuing repeated RDMA operations earlier. 
To satisfy condition 2, memory-side offloading threads return the global address of the updated node upon success (or return failure status on encountering any SMO). 
After an offloading request returns, the compute-side worker thread checks the mapping table to determine if there are any cached copies of the updated nodes and if so, invalidates them by removing their references from the mapping table. 
Such invalidation is rare because the subpath from node $N$ to the leaf node has a high probability of not being in the cache, thanks to path-aware caching and cooling delegation.  
Finally, \name unpins $N$'s parent node and removes $N$ from the mapping table, completing this pushdown request.

\section{Index Operations}
\label{sec:op}
Now we describe how \name performs common index operations. 



\textbf{Lookup and Update.}
As described in Algorithm~\ref{algo:lookup}, 
for any index operation, the compute thread initially navigates the cached tree path,  
employing optimistic lock coupling~\cite{olc} for synchronization. 
When reaching the leaf node,
\name also only requires in-memory locks instead of 
RDMA-based locks thanks to logical partitioning. 

\textbf{Insert.} 
Insert operations may cause node splits to accommodate new keys. 
We use an eager split policy that any full index node encountered during the top-down traversal will be immediately split. 
However, 
since the cached shared nodes (e.g., root) crossing partition boundaries
may not be up-to-date, 
we only apply this policy to index nodes dedicated to the current 
compute server.  
Node splits may propagate up to upper level nodes that are shared. 
We determine whether to split those shared nodes or not by considering their 
freshness. 
Consider the case where an index node $N$ within the current partition is full 
and its parent $P$ is shared, 
we acquire the global lock of $P$ and retrieve its up-to-date version from the memory pool. 
DEX proceeds with and completes the current node split if two conditions are met: (1) cached $P$ is up-to-date and (2) $P$ is not full. 
If not, it means the shared nodes in this traversal path should be first refreshed and possibly split.
As a result, we abandon the ongoing node split and trigger a cache refresh from the remote root with immediate node splits for the shared nodes in the path, if they are full. 
Finally, \name will retry its insert operation and possibly continue the abandoned split for $N$.

\textbf{Delete.} 
\added{  
Similar to node splits, merging nodes may also propagate up to shared nodes. 
Such merging operation proceeds as long as the cached shared nodes are up-to-date. 
Otherwise, a cache refresh from the remote root would be triggered. 
}

\textbf{Range Query.}
To maintain simplicity for the pointer unswizzling process, 
DEX does not maintain links between leaf nodes, ensuring that unswizzling only operates on references from the parent node.
Consequently, range scans that span multiple leaf nodes are subdivided into multiple lookups by employing fence keys.
With lightweight concurrency control, the tree traversal of multiple lookups remains efficient.
After the initial lookup, the traversal of subsequent lookups is generally cached in the CPU cache. 
We do not support operation offloading for range queries 
because a range query may require scanning multiple leaf nodes, 
whose computation load is hard to estimate. 
Our evaluation shows that \name solely with caching and 
one-sided verbs already significantly outperforms state-of-the-art indexes. 
The latency of scanning multiple leaf nodes using one-sided verbs 
can be alleviated via prefetching, 
using the remote leaf pointers stored in the last-level inner nodes, 
which is interesting to be explored in the future.  

\section{Evaluation}
\label{sec:eval}
We evaluate and compare \name with two state-of-the-art range indexes:
\sherman~\cite{sherman} and \smart~\cite{smart}. Major results include:
\begin{itemize}[leftmargin=*]\setlength\itemsep{0em}

\item 
\name achieves 2.5--8.2$\times$ and 4.4--9.6$\times$ higher throughput
than baselines for read- and write-intensive workloads, respectively. 


\item \name's superior performance comes from systematically designed
techniques. With logical partitioning, our cache design improves \name's throughput by up to 15$\times$, and
opportunistic offloading for constrained cache attributes to 55\% further
improvement. 


\item We quantify the cost of logical repartitioning, which can finish in $<$2s
even for compute servers with large caches.


\end{itemize}

\subsection{Experimental Setup}
\label{sec:setup}

\textbf{Testbed.}
We conduct experiments in a cluster of four servers. 
Each server has two 20-core Intel Xeon Gold 6242R CPUs clocked at 3.1GHz (each with 35.75MB
of caches), 384GB DRAM (32GB$\times$12) across two sockets, and a 100Gbps
Mellanox ConnectX-5 \ib NIC connected to a 100Gbps \ib switch. 

To increase the scale of our experiments and stress test \name, we follow prior work~\cite{sherman,smart} to configure each machine to act as one compute server and one memory server.\footnote{RDMA between a compute and a memory server co-located on the same physical machine may not go through the switch.  
However, we observe such accesses already exhibit $\sim$90\% of the cross-server latency going through the switch. 
We therefore believe it is a reasonable tradeoff for larger-scale experiments.} 
On each machine, we allocate 36 cores for the compute server and use the remaining 4 cores for the memory server.
\added{Therefore, the compute power ratio between the compute pool and memory pool is 9:1, 
similar to settings used by previous work~\cite{teleport}. 
The exact deployment shape (capabilities of each infrastructure group) is still being actively explored and to the best of our knowledge, 
the industry has not converged on one particular design (e.g., split vs. pool~\cite{ewais_survey}).
We study different degree of memory disaggregation by varying the the CPU core ratio 
from 36:1 to 9:1 in Section~\ref{sec:sensitivity}.
}
We allocate 256MB of DRAM cache in each compute server (i.e., $\sim$8$\%$ of the bulk-loading dataset size in our evaluations, detailed later in this section) and 64GB of DRAM for each memory server. 
In our experiments, \textit{cache size} refers to the amount of DRAM used as the compute-side cache in one compute server. 
Unless explicitly stated, our experiments use all the 144 (36$\times$4) compute-side threads
and 16 (4$\times$4) memory-side threads in the cluster. Each thread is pinned to a physical core. 
Servers run Arch Linux with kernel 6.3.2.
All the code is compiled with GCC 13.1.1 with all the optimizations enabled.

\textbf{Implementation and Parameters.}
We developed \name using C++. For \sherman and \smart, we use the open-sourced
code from their original authors.\footnote{
\url{http://github.com/thustorage/Sherman} and
\url{http://github.com/dmemsys/SMART}.} 
Since Sherman's original lock-free search using versions~\cite{rdmaguide} is not fully correct, we
implemented an RDMA-based optimistic locking for correct
synchronization~\cite{rdmaguide}. 

For fair comparison,
we configure \sherman and \smart with the parameters as recommended by their original papers~\cite{sherman,smart}.
\sherman uses 1KB fixed-size tree nodes, and \smart uses variable node sizes.
For \name, we use 1KB node size. 
We also set \name's cooling map's capacity to 10$\%$ of the cache. 
Each cooling map bucket occupies 64-byte (one cacheline) to include six FIFO slots.
Subtrees from level 0 to $M=3$ are grouped within the same memory servers. 
For logical partitioning, we range-partition the key space such that each compute server owns an equal key range. 

\begin{table}[t]
  \begin{small}
	\centering
	\caption{\textmd{Microbenchmarks used in our experiments.}}
	\begin{tabular}{l|c|c|c|c}
  \hline
		\bf Workload   & Insert    & Lookup    & Update   & Scan    \\
		\hline 
		\hline
		\bf Read-only   & 0    & 100$\%$   & 0   & 0    \\
		\bf Read-intensive  & 0 & 95$\%$    & 5$\%$    & 0     \\ 
		\bf Write-intensive   & 0    & 50$\%$    & 50$\%$   & 0    \\
		\bf \added{Insert-intensive}   & \added{50$\%$}  & \added{50$\%$}  &  \added{0}  & \added{0}    \\
		\bf Scan-intensive   & 5$\%$  & 0   & 0   & 95$\%$    \\
    \hline
	\end{tabular}
	\label{tab:workload}
	\end{small}
\end{table}

\textbf{Benchmarks.}
We stress test the indexes with microbenchmarks modeled after YCSB~\cite{ycsb}. 
Table~\ref{tab:workload} describes our five workloads. 
The range scan is performed by reading 100 records in ascending order from the initial key.
\added{
The scan-intensive workload evaluates range query performance.
}
Unless explicitly specified, we generate skewed keys for all the workloads, following a Zipfian distribution (theta=0.99) which is the default in YCSB. 
For scalability experiments, we also include tests with uniform workloads.

For all runs, we first bulk-load the index with 200 million key-value records, 
execute a warmup phase comprising 10 million workload operations,
and then benchmark 200 million workload operations. 
For any benchmark exceeding 60 seconds, we collect results from the initial 60 seconds.
Unless otherwise specified, we use 8-byte keys and 8-byte values which can be either an inlined payload or a pointer to an actual record. 

\subsection{Performance and Scalability}
\label{sec:scale}
We examine how each index performs and scales with an increasing number of
compute threads under different workloads. As new compute threads are added, we first exhaust the available cores on existing compute servers, before adding a new one. 
We use all four memory servers to store index nodes.
\added{
	Since logical partitioning is also applicable to \sherman and \smart, we enable it for them to 
	better understand its benefits (denoted as \psherman/\psmart). Specifically, we range-partition the key space 
	such that non-shared nodes do not require RDMA-based synchronization. 
}

\begin{figure*}[]
  \centering
  \includegraphics[width=\textwidth]{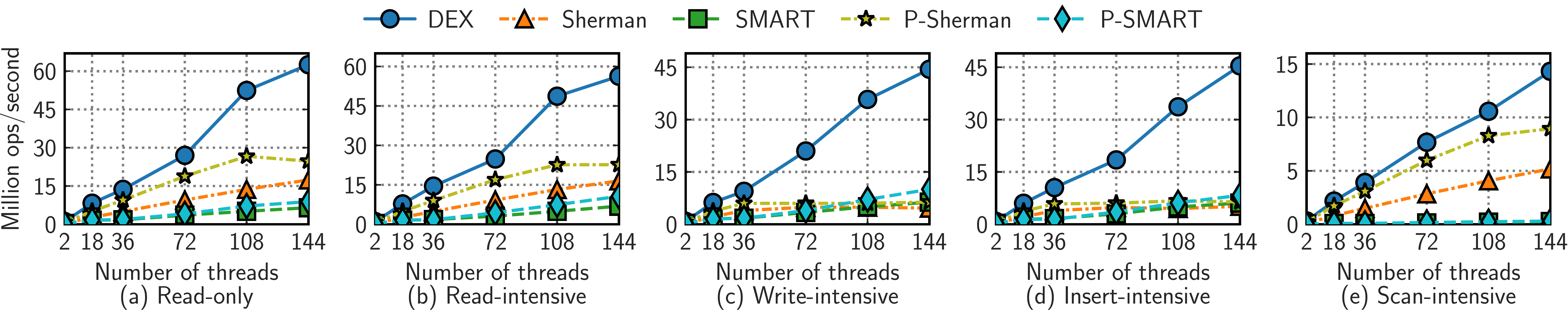}
  \caption{
  \added{\textmd{Throughput under skewed workloads with a varying number of compute threads.}}}
  \label{fig:skew-scalability} 
\end{figure*}

\begin{figure*}[t]
	\centering
	\includegraphics[width=\textwidth]{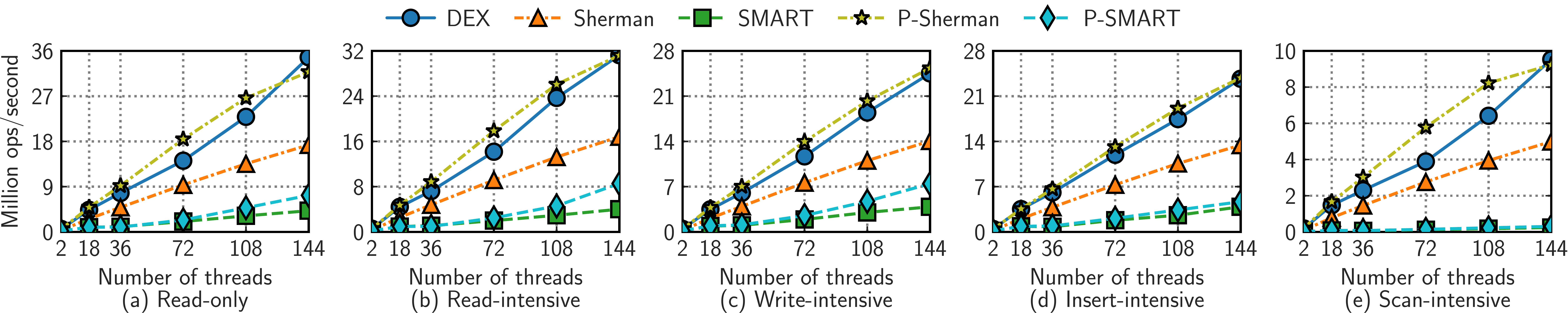}
	\caption{
		\textmd{\added{Throughput under uniform workloads with a varying number of compute threads.}}}
	\label{fig:uniform-scalability} 
  \end{figure*}

  \begin{table}[t]
	\scriptsize
	\begin{small}
	\centering
	\caption{\textmd{RDMA statistics per index operation in skewed read-only (RO)/write-intensive (WI) workloads under 144 threads.}}
	\smallskip\noindent
	\setlength\tabcolsep{1.2mm}
	\begin{tabular}{l|c|c|c|c|c}
  \hline
		\bf Index   & Reads    & Writes    & Atomics   & Two-sided & Traffic (B)   \\\hline
		\hline 
		\bf DEX (RO)   & 0.33  & 0    & 0   & 0.0002  & 333.9  \\ 
		\bf Sherman (RO)   & 3.02    & 0   & 0   & 0 & 1064.69   \\
		\bf SMART (RO)  & 1.44 & 0    & 0    & 0  & 996.99   \\ 
		\bf \added{P-Sherman (RO)}   & \added{1}    & \added{0}  & \added{0}   & \added{0} & \added{1025.04}   \\
		\bf \added{P-SMART (RO)}  & \added{1.15} & \added{0}    & \added{0}    & \added{0}  & \added{397.41}   \\ \hline
		\bf DEX (WI)   & 0.33    & 0.19    & 0   & 0.0001   & 524.1  \\ 
		\bf Sherman (WI)   & 2.71    & 0.99   & 0.59   & 0 & 1078.95   \\
		\bf SMART (WI)  & 1.45 & 0.11    & 0.11    & 0  & 1002.88   \\ 
		\bf \added{P-Sherman (WI)}    &\added{1.02}    & \added{0.5}   & \added{0}   & \added{0} & \added{1054.39}   \\
		\bf \added{P-SMART (WI)}  & \added{1.16} & \added{0.13}    & \added{0}    & \added{0}  & \added{404.207}   \\
    \hline
	\end{tabular}
	\label{tab:skew-rw}
  \end{small}
\end{table}

\textbf{Skewed Workloads.}
As shown in Figure~\ref{fig:skew-scalability}(a), for read-only workloads,
\name scales better and outperforms \sherman/\smart/\added{\psherman/\psmart} 
by 3.6$\times$/9.6$\times$\added{/2.5$\times$/7.1$\times$}, respectively. 
This superiority stems from \name's ability to cache hot tree paths including leaf nodes, with very low coherence overhead among compute servers. 
Conversely, competitors need to retrieve leaf nodes from remote memory through costly RDMA operations, leading to diminished performance.
\added{
\psherman and \psmart exhibit higher performance than \sherman and \smart, respectively, 
thanks to the reduced RDMA-based optimistic reads and better cache locality.
}
Table~\ref{tab:skew-rw} lists the corresponding RDMA statistics for all indexes under 
read-only workloads with 144 compute threads. 
Leveraging efficient caching, 
\name incurs much lower RDMA costs: on average it cuts 
$89\%$/$77\%$/\added{$67\%$/$71\%$} RDMA operations and $69\%$/$67\%$/\added{$67\%$/$16\%$} of RDMA traffic per index operation compared to 
\sherman/\smart/\added{\psherman/\psmart}, respectively. 
Furthermore, we note that \name incurs very few two-sided operations in Table~\ref{tab:skew-rw}. The
reason is that having
sufficient cache capacity in compute servers reduces the need for offloading (more discussions in
Sections~\ref{sec:ablation} and~\ref{sec:sensitivity}). 

As Figure~\ref{fig:skew-scalability}(b) shows, 
\name outperforms \sherman/\smart/\added{\psherman/\psmart} by 3.4$\times$/8.2$\times$/\added{2.5$\times$/5.3$\times$}
under read-intensive workloads. 
SMART exhibits the poorest scalability due to the unscalable FIFO-based caching policy.
Our profiling result shows that its cache admission/eviction takes 
49\% CPU cycles due to severe contention. 
For write-intensive workloads depicted in Figure~\ref{fig:skew-scalability}(c),
\name outperforms \sherman/\smart/\added{\psherman/\psmart} by 9.6$\times$/7$\times$/\added{7$\times$/4.4$\times$}, respectively.
RDMA statistics in Table~\ref{tab:skew-rw} reveal that \name benefits from logical partitioning, avoiding global synchronization overhead (i.e., RDMA atomics) by dedicating each leaf node to a single compute server.
In contrast, \sherman and \smart incur more RDMA atomics and writes, due to 
the manipulation of RDMA-based locks and the immediate write-back of updated leaf nodes to the remote memory pool, hindering their performance and scalability.
SMART marginally outperforms Sherman owing to its write-combining strategy which consolidates multiple concurrent RDMA writes into one.
\added{Although \psherman and \psmart avoid RDMA-based synchornization for leaf nodes, they still lag behind \name due to RDMA traffic of leaf nodes.
The results in insert-intensive workloads depicted in Figure~\ref{fig:skew-scalability}(d) follow the similar trend as write-intensive workloads.}

For scan-intensive workloads in Figure~\ref{fig:skew-scalability}(d),
\name outperforms \sherman/\smart/\added{\psherman/\psmart} by 2.8$\times$/56.3$\times$/\added{$1.6\times$/48.4$\times$},
respectively. SMART and \added{\psmart} lag because each leaf node stores only one key-value
record, necessitating excessive RDMA operations for range scans.

\textbf{Uniform Workloads.}
Most real-world workloads are skewed~\cite{joinskew,queryskew}; 
we use uniform workloads to study the
worst case for caching and \name's performance lower bound. 
  Figure~\ref{fig:uniform-scalability} illustrates that \name consistently
  outperforms \added{\sherman, \smart and \psmart} across all uniform workloads.
  Compared to the results under skewed
  workloads, the performance gap between \name
  and other indexes is smaller.  
  \added{
  Since uniform workloads inherently exhibit much less locality, 
  caching becomes less effective and \name performs similarly to \psherman. 
  Nevertheless, \name can still leverage the limited locality with larger caches, 
  whereas it is impossible for \psherman to do so because it by design does not cache leaf nodes.
  For example, with 512MB cache, DEX improves performance by $\sim$20\%. }

\subsection{Effect of \name Design Choices}
\label{sec:ablation} 
This section quantifies the impact of individual design choices in \name, 
\added{
including the relative contribution of each optimization,} 
effect of our cache design, the cost of logical repartitioning, and the distribution overhead.

\begin{figure}[t]
  \centering
  \includegraphics[width=\columnwidth]{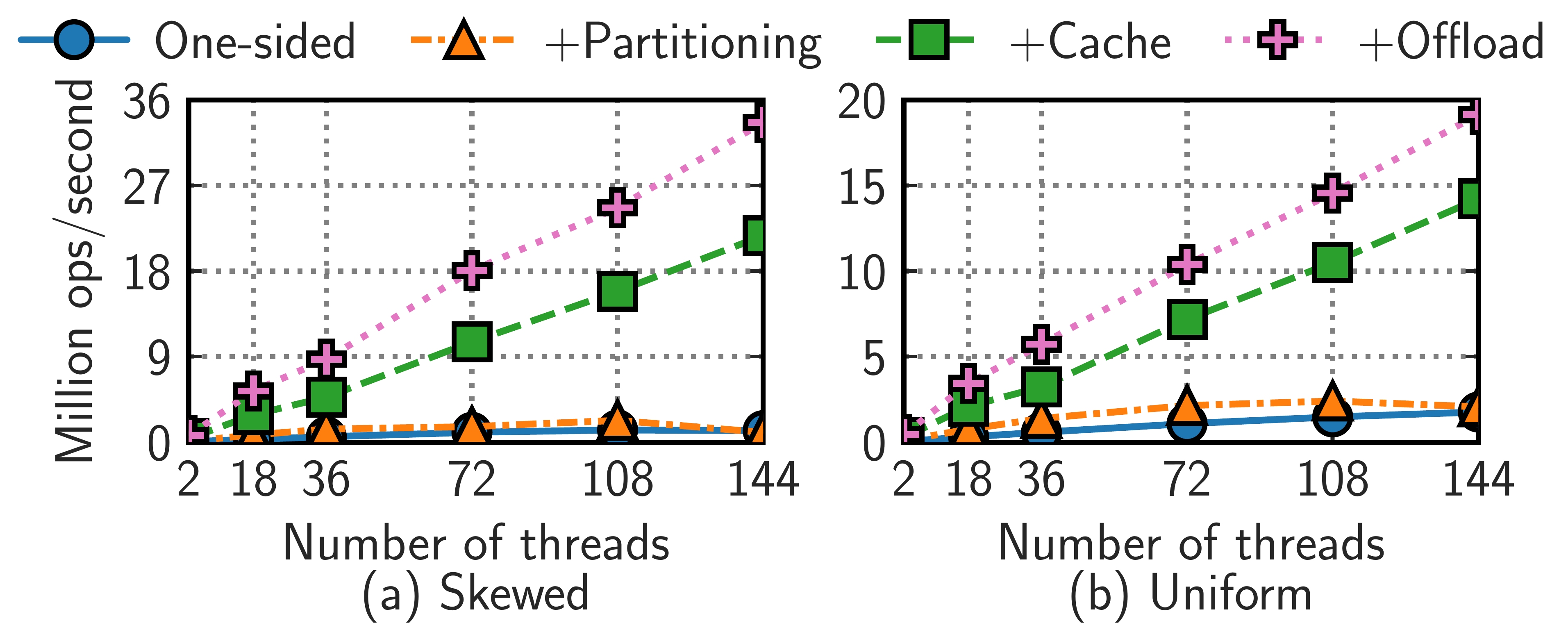}
  \caption{
  \added{\textmd{Effect of each optimization in \name under 31MB cache size and write-intensive workloads.}}} 
  \label{fig:ablation-scalability} 
\end{figure}

\textbf{Ablation Study.}
We study the effect of each optimization under write-intensive workloads. 
Starting from a baseline RDMA B+-tree (described in 
Section~\ref{subsec:tree-on-dm}), 
we add logical partitioning, caching and opportunistic offloading to show 
the throughput improvement under different workloads.
We first enable logical partitioning
because it serves the foundation of our caching and pushdown design.
To trigger offloading, we set the cache size in each
compute server to be much smaller than the working set (e.g., 1\% of the working set or 31MB). 

As Figure~\ref{fig:ablation-scalability}(a) shows,  
the baseline cannot scale under skewed workloads due to excessive remote accesses. 
Under two compute threads (one in each NUMA node), 
logical partitioning improves the throughput
by $2.4\times$ (i.e., from 0.04 to 0.096 Mops/s), thanks to the elimination of RDMA-based synchronization 
on non-shared nodes.
However, with more threads,
network bandwidth becomes the bottleneck again, and the speedup is limited.
Figure~\ref{fig:ablation-scalability}(b) shows similar observations under uniform workloads. 
Further adding caching improves throughput by $21.2\times$/$6.9\times$ under skewed/uniform workloads 
because hot tree paths are cached. 
Finally, adding offloading increases throughput by 55\%/34\% under skewed/uniform workloads, 
benefiting from near-data processing.

\begin{figure}[t]
  \centering
  \includegraphics[width=\columnwidth]{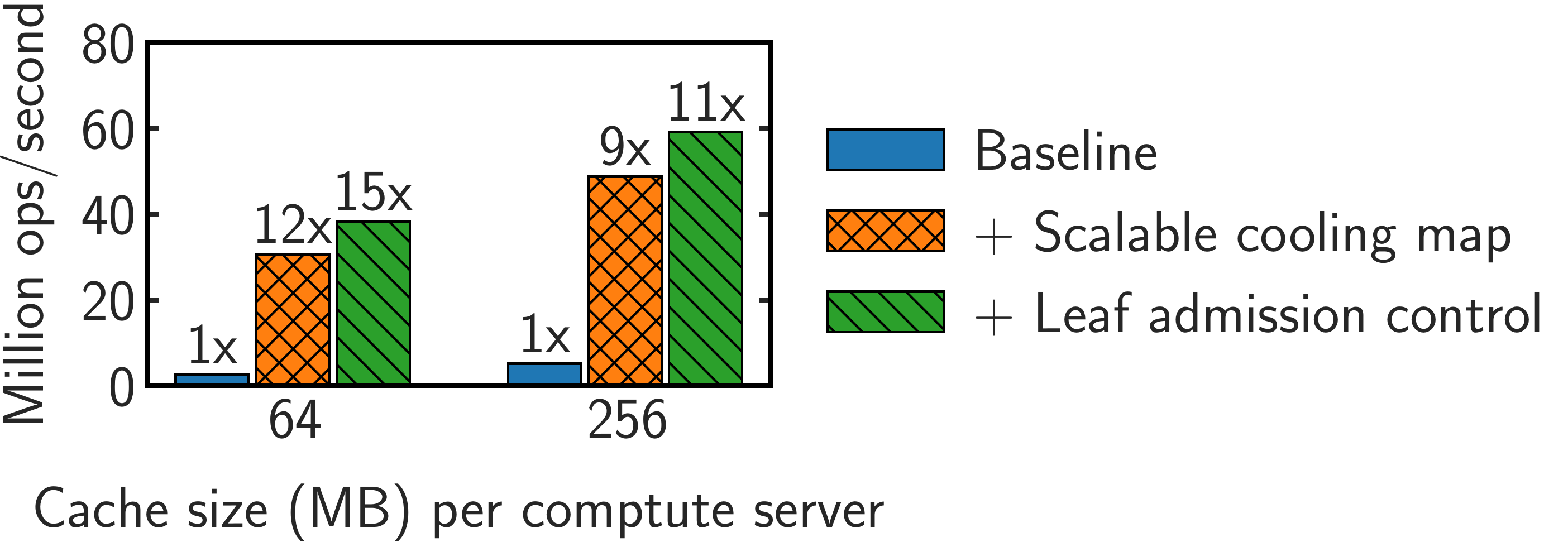}
  \caption{\textmd{Effect of cache design choices in skewed read-intensive workloads under different cache sizes.}}
  \label{fig:ablation-cache} 
\end{figure}

\textbf{Cache Design.}
We evaluate the effectiveness of the cooling map and leaf admission control.
We disable opportunistic offloading here to isolate the effect of these two design choices. 
We experiment with two cache sizes: 64MB and 256MB. The former stresses our replacement algorithm, given its higher cache replacement frequency; 
the latter is the default size.
Starting with a baseline that (1) uses a single lock to protect the cooling map and (2) employs an eager cache admission policy, 
we incrementally introduce other features and measure throughput under 144 compute threads.

The results are shown in Figure~\ref{fig:ablation-cache}. 
Compared to the baseline, using the
cooling map improves \name's throughput by $12\times$ and $10\times$ for
64MB and 256MB cache, respectively. This improvement stems from the use of
fine-grained bucket-level locks in the cooling map, significantly reducing
contention upon cache replacement. On top of that, adding leaf admission control
enables \name to filter out potentially cold pages. This further improves throughput
by $25\%$ and $21\%$ for 64MB and 256MB cache, respectively. This experiment
underscores the efficacy of our caching design, even in scenarios with highly
constrained caches.

\begin{figure}[t]
  \centering
  \includegraphics[width=\columnwidth]{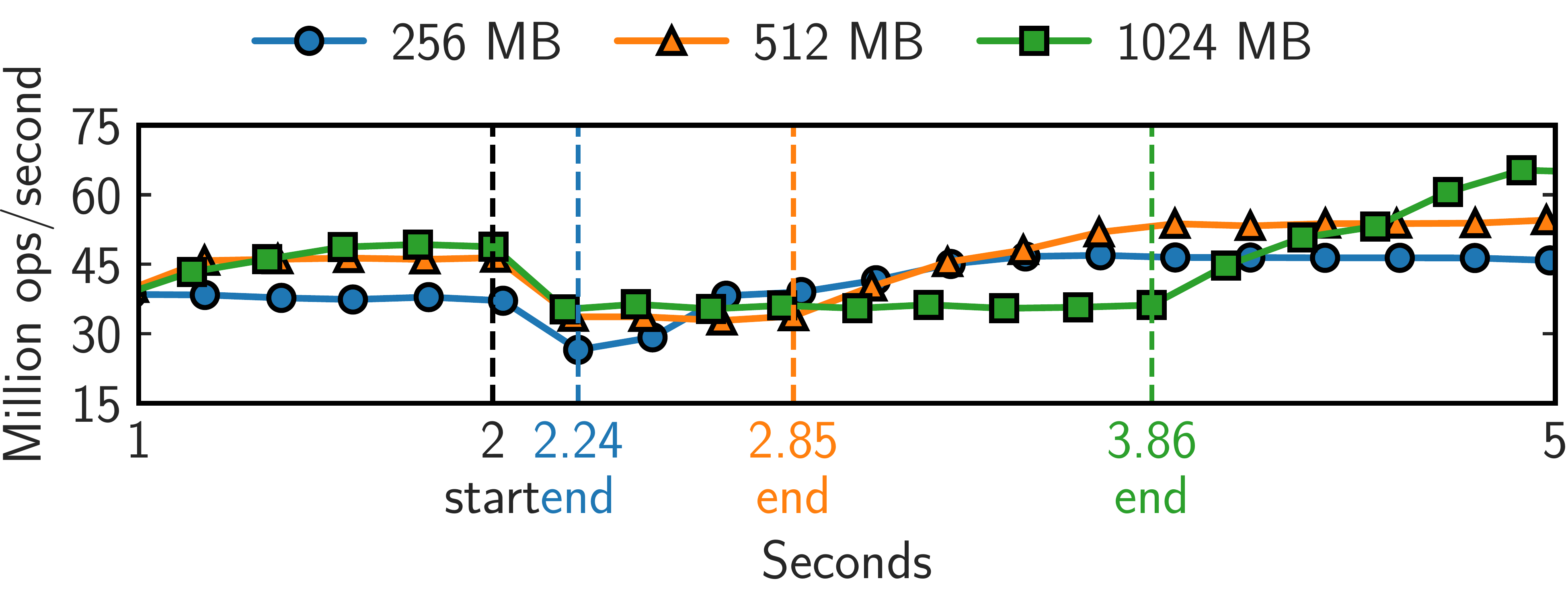}
  \caption{\textmd{Throughput changes during logical repartitioning (started at second 2) under skewed write-intensive workloads.}} 
  \label{fig:repartition} 
\end{figure}
\textbf{Cost of Logical Repartitioning.}
As Section~\ref{sec:partition} describes, \name can promptly repartition to
satisfy the scaling requirement or resolve load imbalance on compute servers. 
We demonstrate this point by observing \name's throughput
changes over time during the repartitioning process. We start with
write-intensive workloads across three compute servers, and then select one
for repartitioning. For a compute server, the cost of repartitioning includes
(1) flushing its dirty cache to the memory pool, and
(2) transferring a portion of its key range to another compute server. 

As shown in Figure~\ref{fig:repartition}, repartitioning begins after the benchmark has run for two seconds. 
\name completes repartitioning within two seconds for cache sizes ranging from 256MB to
1024MB, with larger caches requiring longer time to flush 
dirty cache pages. Notably, these results are based on a single compute thread
for dirty cache flushing; 
employing more compute threads could further
accelerate the repartitioning process. 
After repartitioning, both the repartitioned compute
server and the scale-out (new) server undergo cache warm-up, gradually ramping up the
throughput to normal levels.

\begin{figure}[t]
  \centering
  \includegraphics[width=\columnwidth]{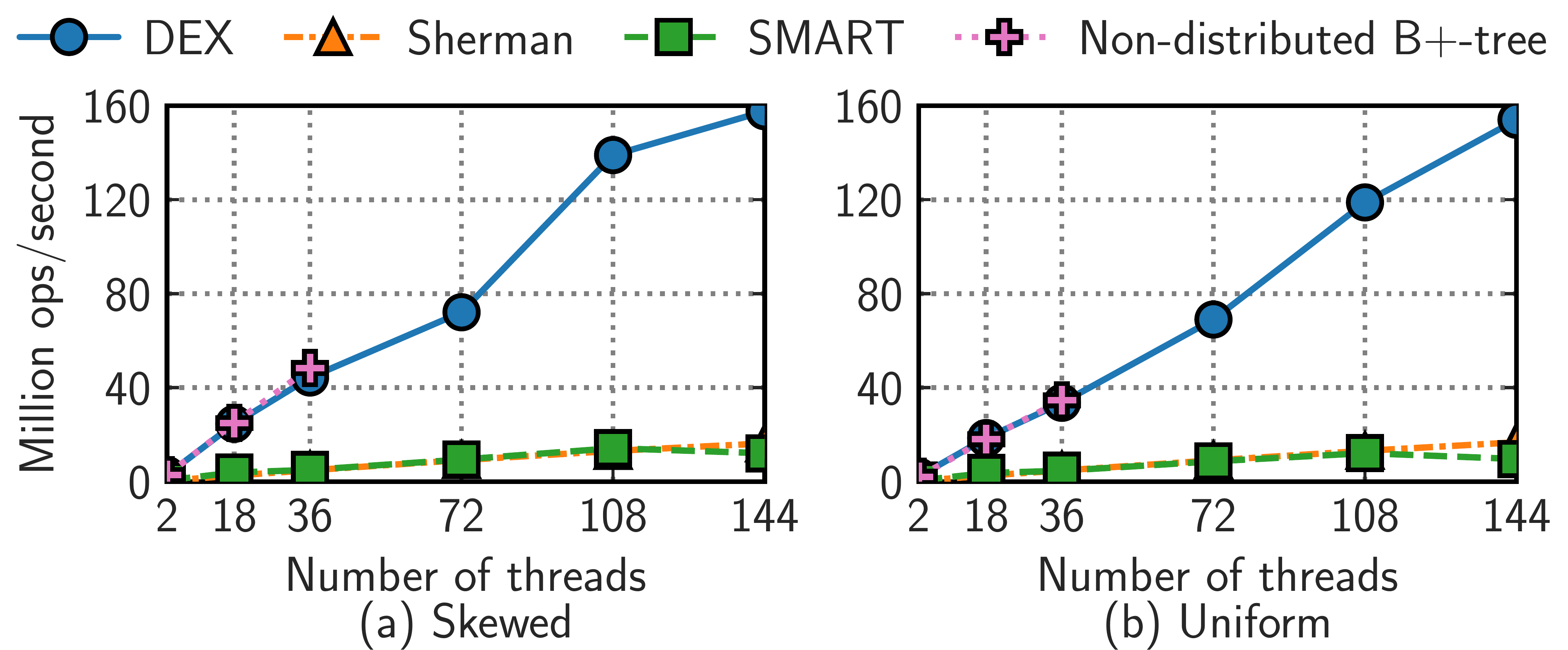}
  \caption{
  \added{Non-distributed B+-tree \textit{vs.} disaggregated memory based tree indexes on read-intensive workloads.}}
  \label{fig:scalability-btree} 
\end{figure}

\textbf{Comparison with Non-distributed B+-Tree.}
To better understand the distribution overhead added by disaggregated memory, we
compare {\name} with a non-distributed in-memory B+-tree running on a single
machine (capable of running up to 36 threads). Our experiments start by ensuring
that both {\name} and the non-distributed B+-tree have the \textit{same} size of
total memory on compute servers. This is fixed to the minimum requirement of
non-distributed B+-tree, or 4.96~GB. We then set {\name} compute servers' caches
for an even distribution --- considering the example of having two compute
servers, each would have a cache of $\frac{4.98}{2} =$ 2.35~GB.

Empirical results are presented in Figure~\ref{fig:scalability-btree}. 
For skewed read-intensive workloads (Figure~\ref{fig:scalability-btree}(a)), 
under 36 threads in a single compute server, 
\name exhibits only a slight performance decrease compared to the non-distributed B+-tree, 
by 8$\%$. This is attributed to the efficient caching 
mechanism of \name, which triggers only approximately 0.01 RDMA verbs per index operation.

Furthermore, Figure~\ref{fig:scalability-btree} highlights the fundamental
trade-off of the non-distributed B+-tree --- being non-scalable, it is limited to
\textit{one} compute server, or up to 36 threads in our experiments. On the other hand, as {\name}
scales out to 144 threads (while reducing each compute server's cache from 4.96~GB
to 1.24~GB), it outperforms the non-distributed B+-tree, with up to 3.3$\times$
higher throughput for skewed read-intensive workloads. We also observe this superior
performance for uniform workloads (Figure~\ref{fig:scalability-btree}(b)), with up to 4.5$\times$ higher throughput. 
It is worth noting that the distribution overhead can significantly vary
for different index designs. An example is how {\sherman} and {\smart} omit
caching the index leaf nodes, leading to a much lower throughput than {\name}.
\subsection{Sensitivity Analysis}
\label{sec:sensitivity}
Now we study how different cache sizes and memory-side compute power
impact index performance.

\begin{figure}[t]
  \centering
  \includegraphics[width=\columnwidth]{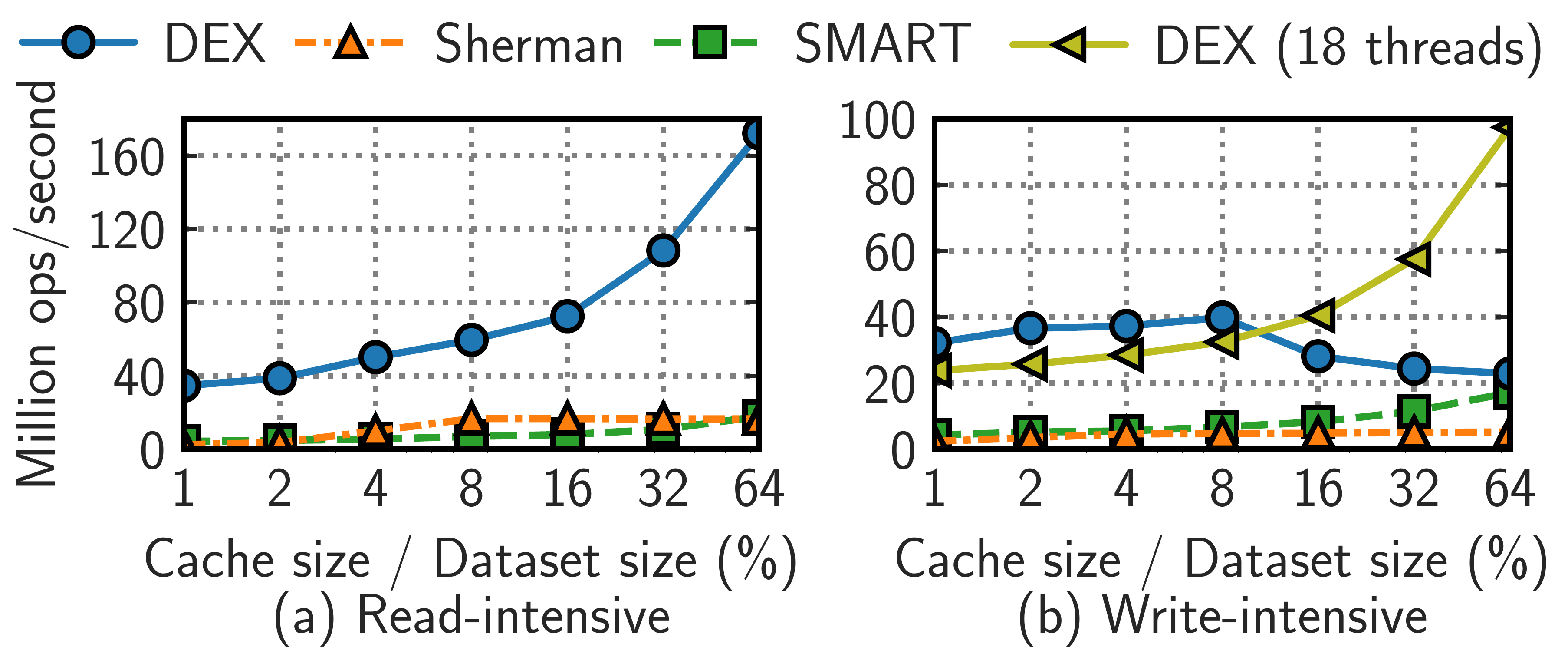}
  \caption{\textmd{Throughput under different cache sizes.}}
  \label{fig:var-cache} 
\end{figure}
\textbf{Cache Size.}
Varying cache size changes cache ratio ($\frac{cache\ size}{dataset\ size}$). 
Figure~\ref{fig:var-cache}(a) shows that \name's performance significantly
improves as the cache ratio increases under skewed read-intensive workloads. Having
a sufficiently large cache enables \name to cache more tree paths, thus incurring fewer 
remote accesses. In contrast, \sherman and \smart do not exhibit the same
benefit from large caches, as they do not cache leaf nodes at all. 
Figure~\ref{fig:var-cache}(b) highlights \name's throughput in skewed write-intensive
workloads under different cache ratios. Its performance improves as the cache ratio
increases from $1\%$ to $8\%$ because of reduced remote accesses. Interestingly, using larger caches 
(cache ratio > 8\%) lowers performance. The reason is that
local synchronization (using optimistic locking) in the cache becomes a
scalability bottleneck since the workload is skewed. 
This becomes particularly severe when we use more than one NUMA node. We verified
this by re-running the experiment with only 18 compute threads pinned to the same socket (labeled
as \textsf{\name(18threads)} in the figure). 
With 18 threads all in one socket, \name scales well without cross-NUMA synchronization. 
We observe the culprit is that the optimistic lock used here is based on centralized spinlocks that are known to be vulnerable to high contention. 
We leave it as future work to address this issue using more recent robust optimistic locks~\cite{optiql} in disaggregated B+-trees. 

\begin{figure}[t]
  \centering
  \includegraphics[width=\columnwidth]{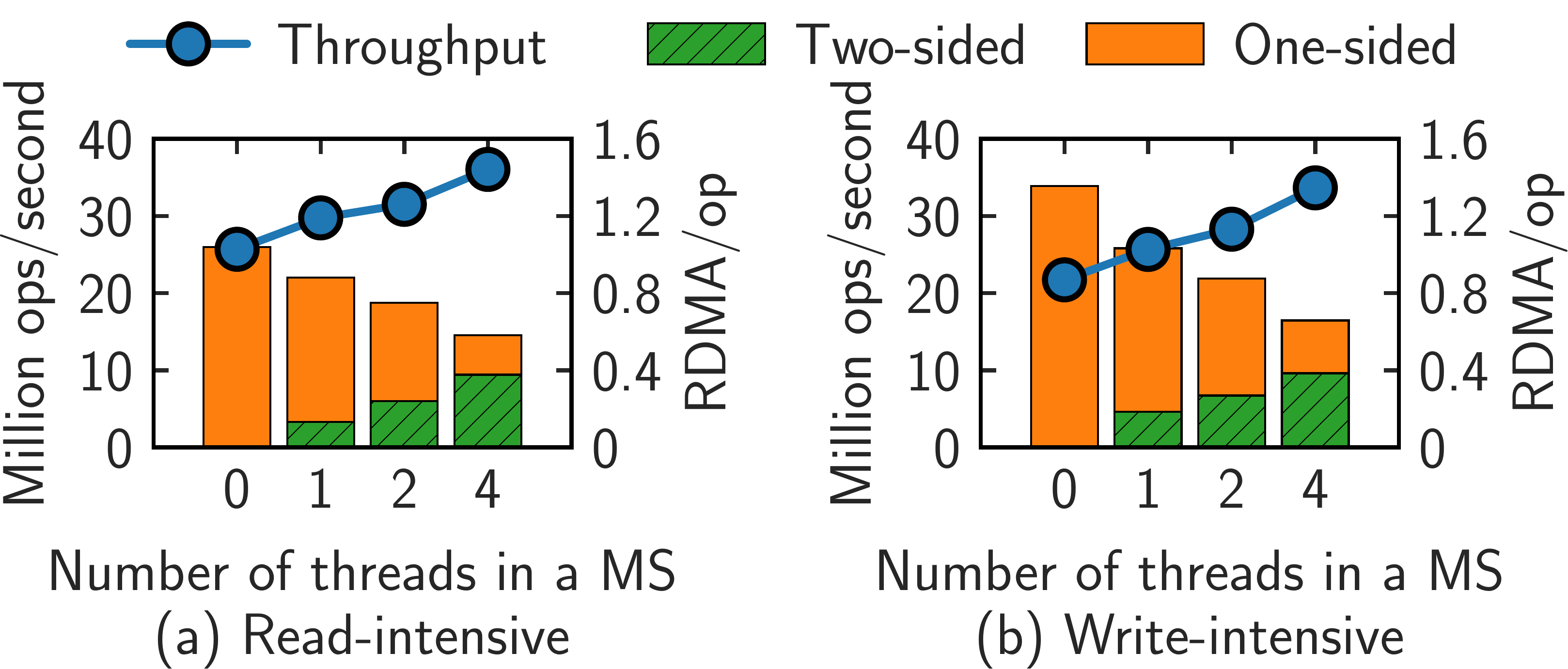}
  \caption{\textmd{Effect of opportunistic offloading under skewed workloads with varying threads in each 
  memory server (MS).}}
  \label{fig:pushdown} 
\end{figure}

\move{
\textbf{Impact of Memory-Side Compute Power.}
Assessing the effectiveness of opportunistic offloading involves varying the number
of memory-side threads serving offloading requests. To trigger offloading, 
as done in Section~\ref{sec:ablation}, we set the cache size in each compute server to
$1\%$ (i.e., 31MB) of the data size.
Figure~\ref{fig:pushdown} shows the throughput and RDMA statistics
under 144 computing threads. As we use more threads to serve offloading requests, 
the number of
RDMA operations including both one-sided and two-sided (incurred
by offloading) is reduced by $56\%$/$49\%$ in read-intensive/write-intensive workloads.
This then leads to a throughput increase of $40\%$/$55\%$. 
As more compute power becomes available in memory servers, 
\name can dynamically
offload more index operations to the memory pool, effectively reducing overall
RDMA costs.}

\section{Related Work}
\label{sec:relateD}

\textbf{Range Indexes for Disaggregated Memory.}
Most DM-optimized indexes are shared-everything. 
FG~\cite{rdmaindex} is the first DM-based B+-tree that
entirely relies on one-sided RDMA. 
\sherman~\cite{sherman} and \smart~\cite{smart} cache inner nodes of tree indexes and 
necessitate remote accesses to leaf nodes. 
dLSM~\cite{dLSM} is a DM-optimized log-structured merge tree 
which adopts a shared-nothing architecture (physical sharding).
\name takes a different approach that is based on logical partitioning for reduced consistency overhead. 

\textbf{Modern Database Caching.}
While \name's cache design is inspired by sampling-based caching approaches, we study and optimize it
for scalable range indexing on disaggregated memory.
LeanStore~\cite{leanstore} 
randomly samples pages and puts them into a shared FIFO list for cooling but exhibit severe contention 
on disaggregated memory. 
A more recent improvement 
is a simpler second-chance strategy~\cite{lean-evolution} where pages already sampled two times are immediately evicted.
It requires dedicated background threads for sampling and timely free-page provision. 
Other sampling-based caching approaches~\cite{watt,scalestore} rank page hotness using epoch information embedded in each cache page and 
also use page-provider threads to avoid stalling worker threads. 
\name uses a cooling map for hotness ranking and scalable eviction on disaggregated memory, without employing dedicated threads.

\section{Summary}
\label{sec:summary}
Disaggregated memory poses unique challenges for building scalable range indexes. 
We observe that achieving high scalability requires a holistic design to 
efficiently utilize limited memory(compute) in compute(memory) pool 
with low consistency overhead.
We present \name, which systematically combines three techniques to reduce 
remote memory accesses and maintain good scalability, as demonstrated through extensive evaluations.

\begin{acks}
  We thank the anonymous reviewers for their constructive comments. 
  We extend our thanks to Fan Yang (Microsoft Research) for his insightful suggestions on paper writing, 
  and to Qianxi Zhang and Ran Shu from Microsoft Research for their extensive discussions during the early stages of this work.
  This work is partially supported by Hong Kong General
  Research Fund (14208023), Hong Kong AoE/P-404/18, 
  the Center for Perceptual and Interactive Intelligence (CPII) Ltd under 
  InnoHK supported by the Innovation and Technology Commission, an NSERC Discovery Grant, 
  Canada Foundation for Innovation John R. Evans Leaders Fund and the B.C. Knowledge Development Fund.
\end{acks}

\bibliographystyle{ACM-Reference-Format}
\bibliography{references}

\appendix
\newpage

\section*{Appendix}

\begin{figure}[t]
  \centering
  \includegraphics[width=\columnwidth]{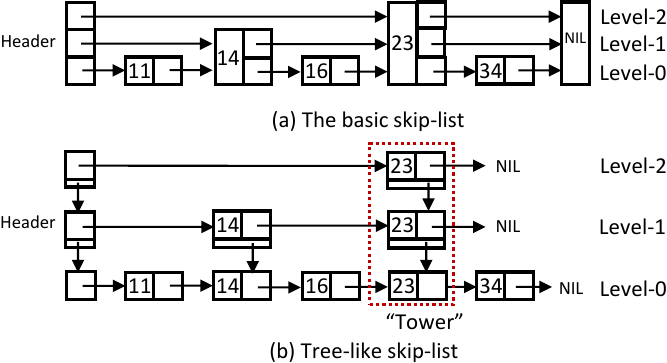}
  \caption{\added{Basic skip-list \textit{vs.} tree-like skip-list.}}
  \label{fig:skiplist-appendix} 
\end{figure}

\section{Generalization to Other Indexes}
\label{sec:generalize}
We discuss the generalization of {\name} beyond B+-trees.
\name techniques are applicable to indexes that satisfy two requirements: 
(1) The index should be a tree-like structure with parent-child relationships, and
the key range of a child node is within the range of its parent node. 
This is necessary for logical partitioning, path caching and offloading. 
(2) The index should have a clear separation between
the index and data layers where the former only guides traffic to the latter without actually storing data. 
This ensures that the data layer would not have
coherence issues due to logical partitioning, and the index layer would require
only lazy cache refresh.

As concrete examples, below we describe how \name can be applied to three 
popular types of indexes: (1) CSB+-tree~\cite{csbtree}, (2) ART (adaptive radix tree)~\cite{ART}, and (3) skip lists~\cite{SkipList}.
However, we note structures like B-trees (which can store data in the index layer) do not satisfy the requirements. 

\textit{(1) CSB+-trees.}
The CSB+-tree aims to optimize CPU cache behavior. 
It is a B+-tree variant that clusters CPU cacheline-sized nodes that are children of the same parent node in a node group. 
Each node group is stored in contiguous memory, and the parent node only stores a pointer to the node group, instead of each individual child.  
Despite these changes, CSB+-tree still satisfies the requirements and applying {\name} is straightforward: 
we simply set the cache page size/cache unit as the node group instead of individual B+-tree nodes.

\textit{(2) ART.}
ART~\cite{ART} is a trie structure that uses four node layouts depending on the number of non-null child pointers. 
Like B+-trees, ART do not store data in inner nodes and 
meets the two requirements.  
However, ART uses multiple node sizes. 
This requires a cache manager that also supports variable-size pages. 
Our current implementation assumes fixed-size pages, but can be extended to support variable-size pages (e.g., with ideas from recent work~\cite{umbra}), which we leave as promising future work.

\textit{(3) Skip lists.}
A skip list is a probabilistic data structure where each level is a linked list of nodes and forward pointers, 
in which the bottom linked list stores data and other layers store guiding information. 
However, 
in the original design~\cite{SkipList} shown in Figure~\ref{fig:skiplist-appendix}(a), 
a node can span multiple levels (e.g., node 23 spans three levels) containing both the guiding information and data,
which violates the second requirement above for generalization. 
In contrast, as Figure~\ref{fig:skiplist-appendix} (b) shows, more recent adaptations~\cite{freeskip} use a tree-like structure that makes \name applicable. 
Specifically, for the same key, there is a separate node at each level. 
These nodes are connected vertically, named a tower. 
In this design, the index layer and the data layer are physically separated: as shown in Figure~\ref{fig:skiplist-appendix}(b), 
levels 1 and 2 compose the index layer, while level 0 is the data layer.

The key range of a node can be determined using its key as the lower bound and 
key of the successor at the same level as the upper bound. 
For example, the key range of node 14 in level-1 is $[14, 23)$, such that node 14 and node 16 in level-0 are the children of it.  
Finally, one caveat is that a skip list node may have multiple incoming pointers such as 
the pointers from the predecessor at the same level and the parent node in a tower. 
Therefore, to evict a cached node, it requires careful designs to unswizzle multiple pointers.

\begin{table}[t]
  \begin{small}
	\centering
	\caption{\added{Additional workload combinations.}}
	\begin{tabular}{l|c|c|c|c}
  \hline
		\bf Workload   & Insert    & Lookup    & Update   & Scan    \\
		\hline 
		\hline
		\bf \added{Read-intensive-2}  & \added{5$\%$} & \added{95$\%$}  & \added{0}  & \added{0}     \\ 
		\bf \move{Insert-only}   & \move{100$\%$}    & \move{0}    & \move{0}   & \move{0}    \\
    \hline
	\end{tabular}
	\label{tab:workload-appendix}
	\end{small}
\end{table}

\begin{figure}[t]
  \centering
  \includegraphics[width=\columnwidth]{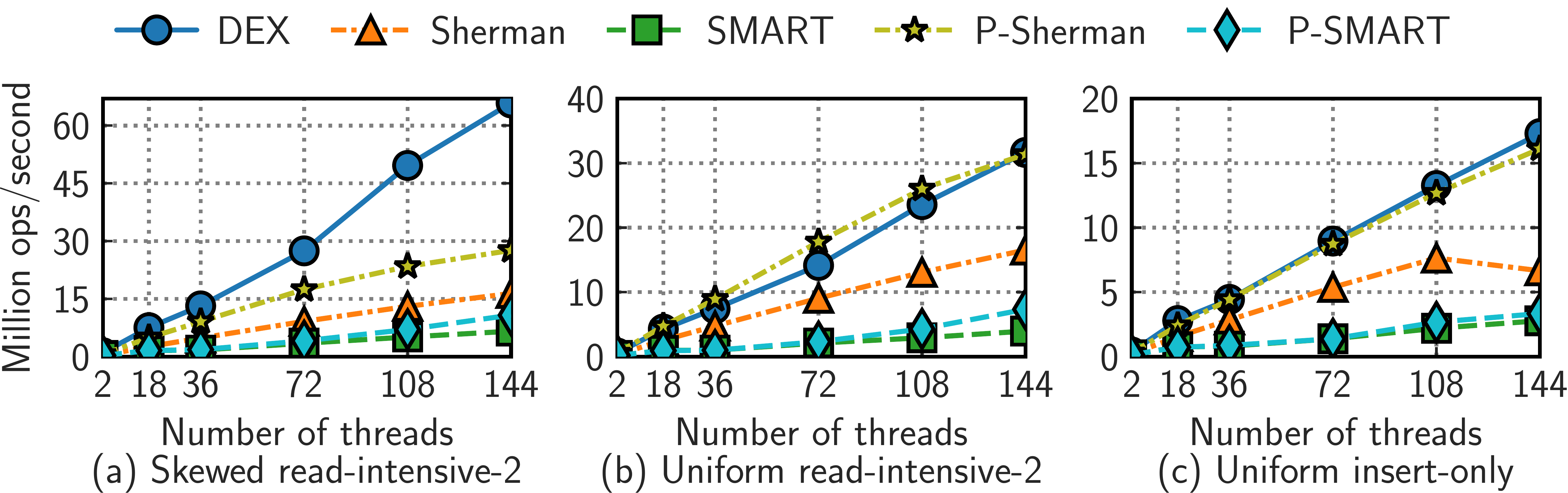}
  \caption{
    \added{
    Throughput under different workloads with a varying number of compute threads.}}
  \label{fig:scalability-appendix} 
\end{figure}

\section{Additional Workloads}
\label{sec:add-workload}
We include two additional workloads in Table~\ref{tab:workload-appendix}.
The read-intensive-2 workload consists of $95\%$ lookup and $5\%$ insert operations.
\move{
The insert-only workload is only tested under uniform key distribution, to guarantee the uniqueness and successful insertion of all keys.
}

As shown in Figure~\ref{fig:scalability-appendix} (a), benefiting from scalable caching of hot tree paths, 
\name outperforms \sherman/\smart/\psherman/\psmart by 4$\times$/10$\times$/2.4$\times$/6.1$\times$ in skewed read-intensive-2 workloads. 
In contrast, under uniform read-intensive-2 workloads, \name achieves competitive performance with \psherman and outperforms others 
with closer performance gap because caching is in nature less effective under uniform workloads.
For uniform insert-only workloads shown in Figure~\ref{fig:scalability-appendix} (c), 
\name achieves the highest throughput among all competitors.

\section{Additional Sensitivity Analysis}
\label{sec:add-sensitivity}
Now we study how different deployment parameters and workload characteristics 
including key size, workload skewness and leaf admission probability
impact index performance.

\begin{figure}[t]
  \centering
  \includegraphics[width=\columnwidth]{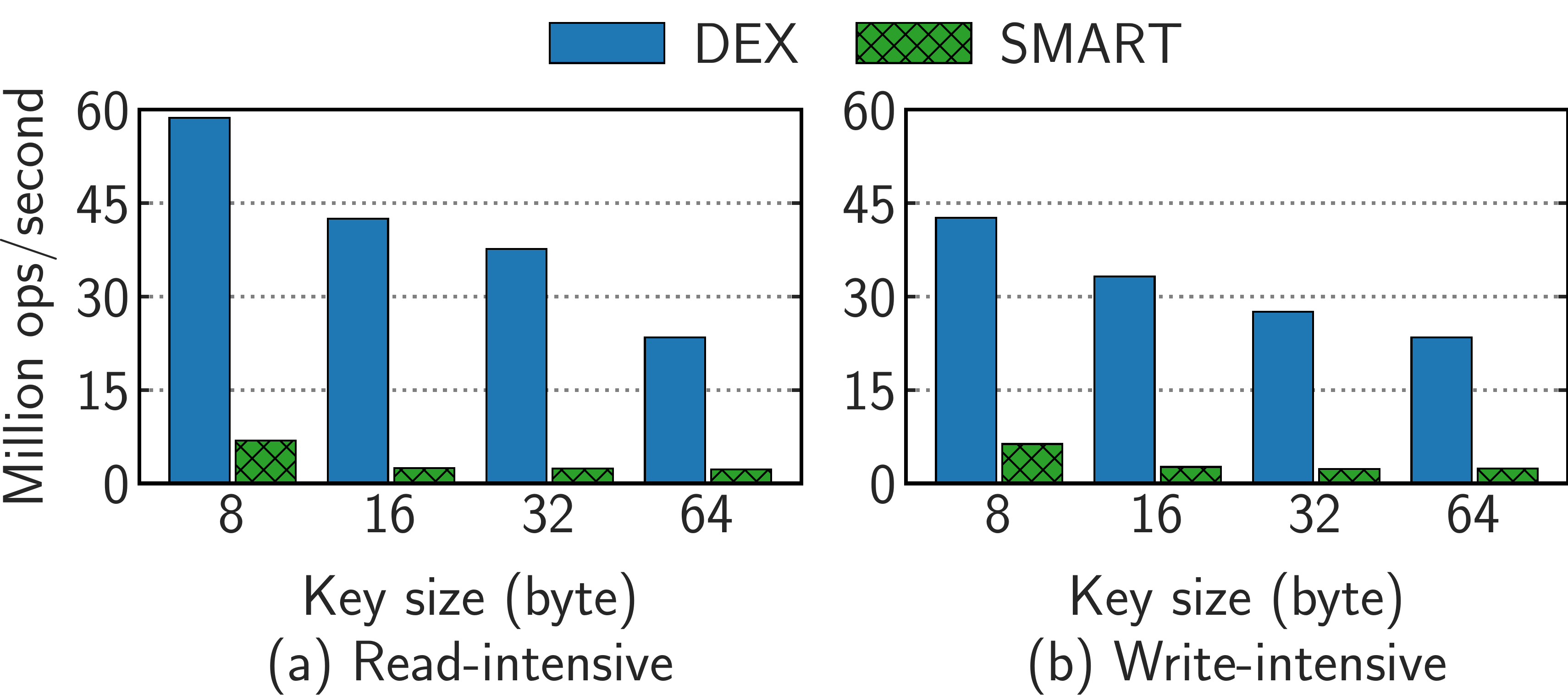}
  \caption{Throughput under skewed workloads and different key sizes.}
  \label{fig:key-size} 
\end{figure}

\textbf{Key Size.}
We evaluate how different key sizes impact the index performance. The keys are
embedded in fixed-sized index nodes (1KB) of \name. We concatenate randomly-generated 8B hashed keys as strings and use them as the datasets for larger key
sizes. Our comparison baseline here is \smart because the open-sourced implementation of
\sherman does not support keys longer than 8 bytes. 
In Figure~\ref{fig:key-size}, as the key size increases, both \name and \smart
exhibit a performance drop in all workloads. The reason is the increased data
size (2.98GB for 8B key, to 13.41GB for 64B key) naturally leads to larger indexes and
stresses the constrained fixed-size cache in compute servers. 
\name still maintains a performance advantage over SMART, thanks to its hybrid
caching and offloading strategies that can effectively reduce remote
accesses.

\begin{figure}[t]
  \centering
  \includegraphics[width=\columnwidth]{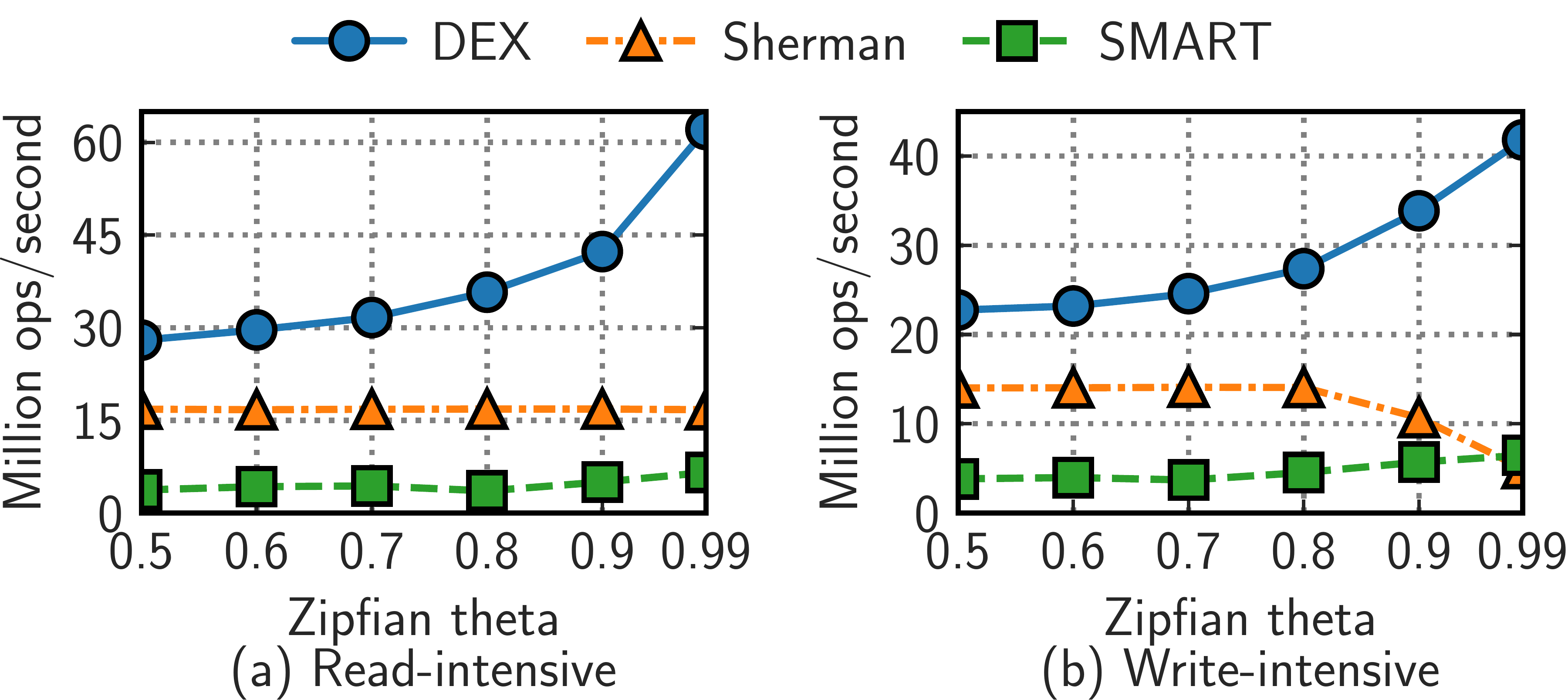}
  \caption{Throughput under varying workload skewness. A larger theta value indicates more skewness. }
  \label{fig:skewness} 
\end{figure}

\textbf{Workload Skewness.}
We evaluate each index under different workload skewness, by varying the
$theta$ parameter of the Zipfian distribution. As shown in
Figure~\ref{fig:skewness}, with a higher theta (skewness), \name performs better because it
mostly accesses a small set of hot keys, enabling more hot tree paths to be
cached in compute servers. This results in fewer remote memory accesses. 
In contrast, \sherman and \smart do not cache any leaf node. Even worse,
under very high skewness ($theta$=0.99) in write-intensive workloads shown in
Figure~\ref{fig:skewness}(b), \sherman's throughput collapses as RDMA-based synchronization now becomes a major bottleneck.

\begin{figure}[t]
  \centering
  \includegraphics[width=0.85\columnwidth]{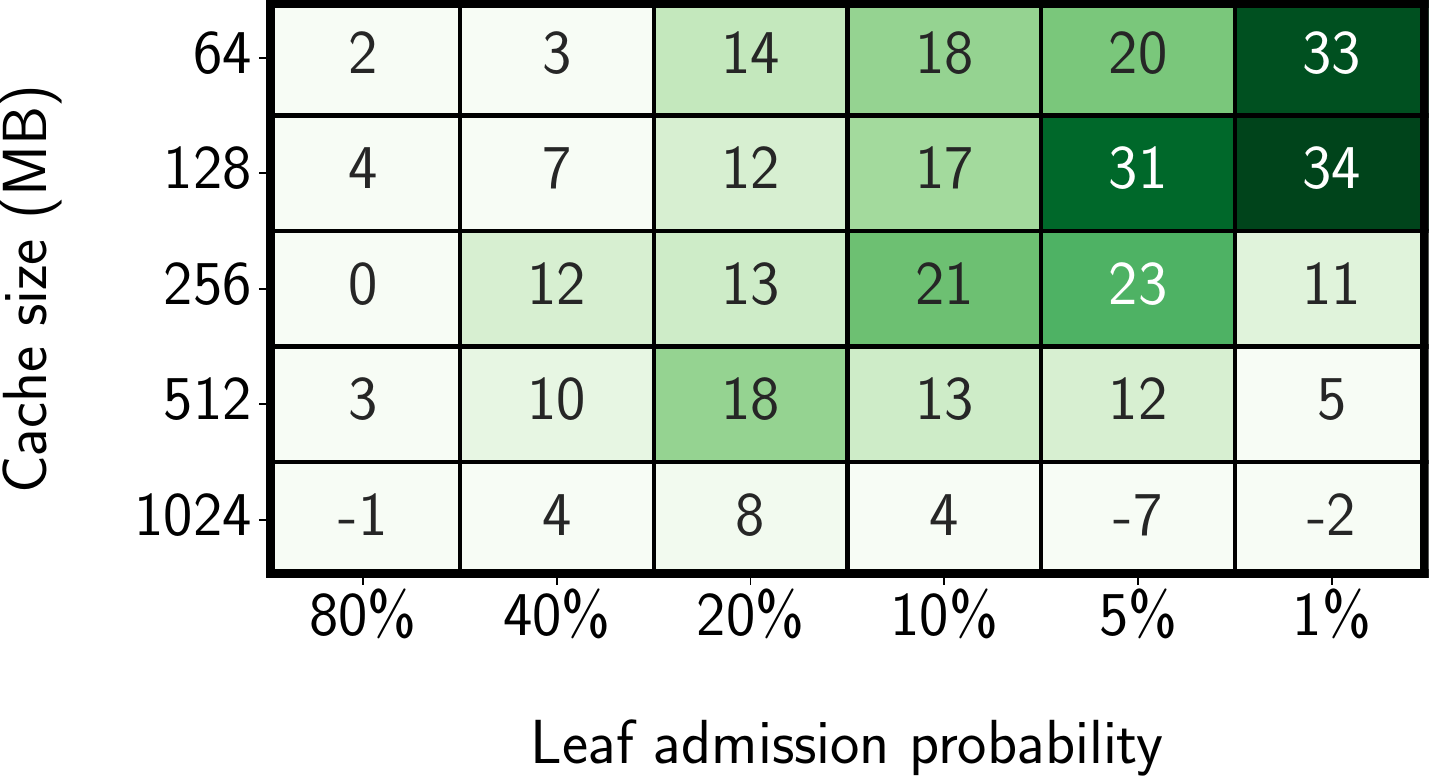}
  \caption{Percentage of \name throughput changes over a baseline that unconditionally admits all accessed nodes ($P_A=100\%$) under different cache sizes and leaf
  admission probabilities, under read-intensive workloads.} 
  \label{fig:pa} 
\end{figure}

\textbf{Leaf Admission Probability ($P_A$).}
$P_A$ controls how lazily (low $P_A$) or eagerly (high $P_A$) \name should admit leaf nodes. 
Intuitively, with a small cache, a lower $P_A$ helps avoid caching potentially cold
pages. Otherwise, a higher $P_A$ is preferred to take advantage of the
relatively ample cache space. To this end, Figure~\ref{fig:pa} evaluates how
$P_A$ impacts throughput, by using $P_A=100\%$ (always admit) and read-intensive workloads
as the baseline. For 64--128MB caches, we observe that a low $P_A$ (e.g.,
1$\%$) helps improve \name's performance by up to $34\%$. When we increase the cache
size to 256--512MB, setting $P_A$ to $5\%$ and $20\%$ becomes the more optimal option. 
As we further increase the cache size to 1024MB, lazy admission becomes too
conservative to improve throughput and can even slow down performance by 7\%.
Therefore, while our current implementation defaults $P_A$ to $10\%$, manually
or automatically tuning $P_A$ might still be necessary to achieve the
best performance for a given deployment, which we leave for future work.


\end{document}